\documentclass[preprintnumbers, aps, floatfix, onecolumn, preprintnumbers, letterpaper, superscriptaddress,nofootinbib,10pt]{revtex4-2}
\usepackage{dcolumn}
\usepackage{graphicx}
\usepackage{amsmath}
\usepackage{amsfonts}
\usepackage{amssymb}
\usepackage{microtype}
\usepackage{subfigure}
\usepackage{makeidx}
\usepackage{bm}
\usepackage{epsf}
\usepackage[colorlinks=true,citecolor=blue,linkcolor=black,urlcolor=black]{hyperref}
\usepackage{color}
\usepackage{multirow,dcolumn}
\usepackage{graphicx}
\usepackage{mathrsfs}
\graphicspath{{Images/}}

\usepackage{mathrsfs}
\usepackage{dcolumn}
\usepackage{graphicx}
\usepackage{amsmath}
\usepackage{amsfonts}
\usepackage{amssymb}
\usepackage{microtype}
\usepackage{subfigure}
\usepackage{makeidx}
\usepackage{bm}
\usepackage{epsf}
\usepackage{color}
\usepackage{multirow,dcolumn}
\usepackage{graphicx}
\usepackage{mathrsfs}
\graphicspath{{Images/}}

\def\doi{http://doi.org}

\def\be{\begin{equation*}}
\def\ee{\end{equation*}}

\def\Ref{\ref}

\begin{document}
\begin{flushleft} 
KCL-PH-TH/2023-{\bf 02}
\end{flushleft}

\title{Regular Compact Objects with Scalar Hair}

\author{Thanasis Karakasis}
\email{thanasiskarakasis@mail.ntua.gr} \affiliation{Physics Division, School of Applied Mathematical and Physical Sciences, National Technical University of Athens, 15780 Zografou Campus,
Athens, Greece.}

\author{Nick E. Mavromatos}
\email{mavroman@mail.ntua.gr} \affiliation{Physics Division, School of Applied Mathematical and Physical Sciences, National Technical University of Athens, 15780 Zografou Campus,
Athens, Greece.}
\affiliation{Theoretical Particle Physics and Cosmology group, Department of Physics, King's College London, London WC2R 2LS, UK.}

\author{Eleftherios Papantonopoulos }
\email{lpapa@central.ntua.gr} \affiliation{Physics Division, School of Applied Mathematical and Physical Sciences, National Technical University of Athens, 15780 Zografou Campus,
Athens, Greece.}

\begin{abstract}

We discuss  exact regular compact object solutions in higher dimensional extensions of General Relativity sourced by a phantom scalar field in arbitrary $D$ spacetime dimensions ($D>2$), for which a central singularity is absent.  We follow a bottom-up approach, by means of which, by imposing the desired form of the solution to the metric function,  we derive the form of the self-interaction scalar potential, which in general appears to depend on both the scalar-hair charge and the black-hole mass.  We discuss in this context the validity of the first law of thermodynamics in such systems.
Consistency  requires the independence of the potential of the mass, imposing in this way the dependence of the mass on the scalar charge of a type that varies with the value of $D$, and according to the no-hair theorem dressing the regular black hole solution with secondary hair. In $D=3,4$ we demonstrate that the potential depends  on the ratio of the scalar charge over the mass, and thus considered as a parameter of the theory.  This feature, however, does not characterise higher-dimensional cases.  Calculating the $D-$dimensional Kretschmann scalar  we show that it is finite at the centre point $r=0$ for arbitrary $D$, rendering the solutions regular. The phantom matter content of the theory is also regular at $r=0$, hence the radial coordinate of our manifold is defined for $r \ge 0$.  We explicitly discuss  the cases of $D=3,4,5,6,10$, and demonstrate that we can have regular, asymptotically flat, black holes with secondary scalar hair.
\end{abstract}

\maketitle

\tableofcontents

\section{Introduction}

In string theory, and its brane extensions~\cite{string}, it is well known  that we need more than four dimensions in order to describe the general structure of the Universe, in the sense of constructing a consistent and unified quantum theory of the fundamental forces observed in nature, including gravity. This idea was first put forward in the context of the extra-dimensional Kaluza-Klein theory~\cite{Kaluza,Klein}, which is known to provide the starting point of string theory. The existence of extra dimensions in space, compactified either at every spatial point  of the four-dimensional spacetime by means of appropriate compact extra-dimensional spatial manifolds, such as Calabi-Yau spaces, or via appropriate boundary conditions on appropriate branes in the extra-dimensional bulk space, opens up new possibilities for interpreting the dark sector of the Universe, by means of fields living in the extra dimensions.

The extra dimensions may exhibit a
dynamical behavior, by varying with the cosmic time. In the literature there are many works studying the cosmological implications of the extra dimensions, under the assumption
 that the observed universe emerges from higher dimensions. For instance, to describe the early inflationary era of the four-dimensional observed universe,  it is often assumed that compactification of extra dimensions happened before the inflation phase  which indicates that the compactification condition is considered during the birth of the Universe. In \cite{Mohammedi:2002tv} a cosmological model based on Kaluza-Klein theory was studied. The Freedmann-Robertson-Walker equations of standard four dimensional cosmology were obtained precisely. The pressure in this Universe was obtained as an effective pressure expressed in terms of the components of the higher-dimensional energy-momentum tensor and  a cosmic scale factor was contracted from the extra dimensional contributions which must be positive providing  a compactification condition.

In standard cosmology after creation of matter at the end of inflation, the Universe enters a deceleration regime and then, at late epochs, returns to an accelerated phase (the quintessence epoch) again. Therefore the  formation  of the extra dimensions may give more information in understanding dark energy or the large-scale structure of the Universe as a whole. In \cite{Ketov:2017aau}, in  a higher-dimensional gravity theory in the presence of a cosmological constant, after performing a conformal transformation, an inflationary scalar potential was obtained by making use of the compactification condition, leading to the Starobinsky inflationary model~\cite{Starobinsky:1980te}. In \cite{Keskin:2019mve} in a $f(R)$ gravity theory a D-dimensional Kaluza-Klein type theory was studied. It was shown that  additional
dimensions are compactified on the standard four spacetime dimensions in the deceleration phase speculating that
the additional dimensions may be responsible for the dynamics of the dark energy dominated era indicating that dark energy may be the source of extra dimensions.

The recent observational results on the other hand indicate that in the early-time cosmological evolution the matter structure formation was
governed by a peculiar form of a cosmic fluid, the dark energy of which is characterized by negative values of the  pressure to density ratio $w$ with $w < -1$ \cite{steinhardt,tegmark,seljak,hannestad,star03,chandra}  and also negative energy density. This kind of dark energy in order to give negative value of $w$ should be parametrized by a {\it phantom field} having negative kinetic energy \cite{sen,gorini,fara05}. However, in that case, a perfect-fluid description of dark energy is plagued with instabilities at small scales due to an imaginary velocity of sound that characterise the phantom-matter case. To avoid this  instability, a phantom scalar field may  be regarded as an effective field description following from an underlying theory with positive energies \cite{no03,trod}.

In a recent paper \cite{Farrah:2023opk} it was claimed that regular black holes (BHs) can be supported by astrophysical and cosmological observations as  realistic astrophysical BH models which can become cosmological at large distance from the BH. In this way non-singular cosmological BH models can couple to the expansion of the universe, gaining mass proportional to the scale factor. This claim was based on a recent study of supermassive BHs within elliptical galaxies where it was found preferential BH growth, relative to galaxy stellar mass \cite{akiyama2022first}. This leads to a realistic behavior at infinity of BH models   predicting that the gravitating mass of a BH can increase with the expansion of the universe independently of accretion or mergers, in a manner that depends on the BH interior solution. Then in \cite{Farrah:2023opk} it was proposed that stellar remnant BHs are the astrophysical origin of dark energy, explaining the onset of accelerating expansion of the Universe.

Local solutions of a gravity theory with a scalar field  minimally coupled to gravity with
arbitrary potentials and negative kinetic energy were investigated in \cite{Bronnikov:2005gm}. It was found that
regular  configurations were formed by the phantom scalar field in  flat, de Sitter and AdS asymptotic spacetimes,
avoiding the BH central singularity. Their main motivation was to find regular BH solutions with an expanding,
asymptotically de Sitter Kantowski-Sachs  cosmology beyond the event horizon. In the literature there are other solutions
alternative to known ones, with a regular center~\cite{dym92,ned01,bdd03}.

Motivated by the above discussion we will study in this work a higher-dimensional gravity theory in the presence of a self-interacting phantom
scalar field minimally coupled to the higher-dimensional gravity.  Singularities are a common feature of BH physics. In most case, the singularities are located at the center of the coordinate system, where curvature invariants possess a divergence and the spacetime is geodesically incomplete. However, Penrose showed that any singularity has to be covered by an event horizon (cosmic censorship hypothesis) and, as a result, all pathologies occurring at the singular region do not affect observers and physics outside of the horizon.

In this work, we shall look for regular balck hole solutions in various dimensions and shall examine their energy conditions. By regular, we mean that there are no gravitational singularities either in the interior or the exterior region of the compact object. Compactness implies that the relevant celestial object possesses an event horizon with a radius smaller than that of the corresponding one in the Schwarzchild geometry. In particular, we consider a phantom scalar field with a scalar charge $A$ in a high-dimensional gravity theory. If the scalar charge is zero then the gravitational singularity is covered by a horizon and then we have a normal BH with a constant scalar field. However, if $A$ is not zero then the scalar charge of the phantom scalar field deforms the geometry in such a way that the gravitational singularity is absent and a compact object, being a regular BH. We  use a specific  form of the metric function to derive the form of the self-interaction scalar potential by solving appropriately the system of the Lagrange equations of motion of the gravitational field theory.

We find that the  charge of the scalar field is connected to the mass of the BH dressing in this way the BH with secondary hair \footnote{We refer the reader to \cite{primary} for some examples of black hole solutions with a primary scalar hair.}. This results to the  independence of the resulting scalar potentials on the compact object's mass, which in turn imposes restriction on the mass, and leads to its dependence on the scalar charge, thus rendering the hair secondary. However, this is possible only for the  dimensionalities $D=3,4$, where the scalar potential depends only on the ratio of the scalar charge over the mass.
We also discuss the first law of thermodynamics, which turns out to be not well understood for regular black holes \cite{Ma:2014qma, Lan:2023cvz}. This happens because in order to build a non-singular black hole, one has to handle the energy momentum tensor appropriately, which in turn implies handling the matter Lagrangian in a particular way and consequently, the Lagrangian depends on the primary black hole charges, such as the mass of the black hole, rendering the identification of the black hole mass as the internal energy of the black hole ill-defined. As a result, one cannot consider that the internal energy of the black hole and the mass coincide. In the case of $D=3$ we are able to perform simple calculations and show that indeed this is the case, while when the secondary hair $A$ vanishes (in which case we obtain the BTZ black hole) the internal energy coincides with the mass of the black hole.

The structure of the article is as follows: in the next section, \ref{sec:buildBH}, we construct a non-singular black-hole-like compact object in a theory with scalar phantom matter fields, and explain in detail the specific violation of the no hair theorem that allows such a hair. In section \ref{sec:cases} we discuss various examples, in different space time dimensions, $D \ge 3.$: we start from the $D=3$ case,  then revise the $D=4$ case of \cite{Bronnikov:2005gm}, discussing thermodynamics, which was not discussed in that work, and then proceed to discuss cases with higher spacetime dimensions. Finally in section \ref{sec:concl} we conclude.

\section{Setup of the theory and $D$-dimensional regular black holes}\label{sec:buildBH}

We consider the action
\begin{equation} S = \int d^Dx \sqrt{-g}\left\{ \frac{R}{2\kappa} - \frac{1}{2}f(\phi)\nabla_{\mu}\phi\nabla^{\mu}\phi - V(\phi)\right\}~,\end{equation}
which consists of the Ricci scalar $R$~\footnote{Our conventions and definitions throughout this paper are: $(-,+,+,+)$ for the signature of the metric, the Riemann tensor is defined as
$R^\lambda_{\,\,\,\,\mu \nu \sigma} = \partial_\nu \, \Gamma^\lambda_{\,\,\mu\sigma} + \Gamma^\rho_{\,\, \mu\sigma} \, \Gamma^\lambda_{\,\, \rho\nu} - (\nu \leftrightarrow \sigma)$,
and the Ricci tensor and scalar are given by  $R_{\mu\nu} = R^\lambda_{\,\,\,\,\mu \lambda \nu}$ and $R= g^{\mu\nu}\, R_{\mu\nu}$ respectively.}. and a self interacting scalar field, minimally coupled to gravity with $\kappa=8\pi G_D$ where we will set $G_D=1$ from now on. In the action, $f(\phi)$ controls the nature of the scalar field, being either a phantom $(f(\phi)<0)$ or a regular one $(f(\phi)>0)$. The field equations read
\begin{eqnarray}
&&G_{\mu\nu} =\kappa T_{\mu\nu} ~,\\
&&f(\phi)\Box\phi +\frac{f'(\phi)}{2}\nabla_{\mu}\phi\nabla^{\mu}\phi  = \frac{dV}{d\phi}~,\\
&&T_{\mu\nu} = f(\phi) \nabla_{\mu}\phi\nabla_{\nu}\phi - \frac{f(\phi)}{2}g_{\mu\nu}\nabla_{\alpha}\phi\nabla^{\alpha}\phi - g_{\mu\nu}V(\phi)~.
\end{eqnarray}
We shall consider the following metric ansatz
\begin{equation} ds^2 = - b(r)dt^2 + \cfrac{1}{b(r)}dr^2 + W(r)^2 d\Omega^2_{D-2} \label{ds}~,\end{equation}
where
\begin{equation} d\Omega^{2}_{D-2} = d\theta_{1}^2 + \sum_{i=2}^{D-2} \prod_{j=1}^{i-1}\sin^{2}\theta_{j}d\theta_{i}^2~. \end{equation}
This ansatz allows us to obtain the Schwarzchiuld metric, as a smooth limit when the phantom scalar field is absent. 
We can calculate the components of the Einstein equation which under this ansatz reduces to the following relations
\begin{eqnarray}
&&(D-2) W''(r)+8\pi  f(r) W(r) \phi '(r)^2=0~, \label{ttrr}\\
&&W(r) \left(2 b(r) W''(r)-(D-4) b'(r) W'(r)\right)+W(r)^2 \left(-b''(r)\right)+2 (D-3) \left(b(r) W'(r)^2-1\right)=0~, \label{ttuu}\\
&&V(r) = -\frac{(D-2) W(r) \left(b'(r) W'(r)+2 b(r) W''(r)\right)+(D-3) (D-2) \left(b(r) W'(r)^2-1\right)+\kappa  b(r) f(r) W(r)^2 \phi '(r)^2}{16 \pi  W(r)^2}~, \label{v}
\end{eqnarray}
from which we can obtain directly the metric function $b(r)$
\begin{align} 
b(r)  = &c_1 W(r)^2-c_2 W(r)^2 \left(\int \frac{1}{W(r)^{D}} \, dr\right) \nonumber \\ &-2 (D-3) W(r)^2 \left\{\left(\int W(r)^{D-4} \, dr\right) \int \frac{1}{W(r)^{D}} \, dr
-\int W(r)^{D-4} \left(\int \frac{1}{W(r)^{D}} \, dr\right) \, dr\right\}~,\end{align}
where $c_1,c_2$ are constants of integration. Setting $W(r)=r$ we obtain
\begin{eqnarray}
&&\phi(r)=\text{constant}~,\\
&&V(r) = -\frac{c_1 (D-2) (D-1)}{16 \pi }~,\\
&&b(r) =1+c_1 r^2+\frac{c_2 r^{3-D}}{D-1}~,
\end{eqnarray}
which is just the family of (A)dS black hole solutions for arbitrary dimension $D$, $c_1$ is related to the cosmological constant and $c_2$ to the mass of the black hole. The Kretshmann scalar $K= R_{\alpha\beta\gamma\delta}R^{\alpha\beta\gamma\delta}$ is singular at the origin for $D>3$ as expected and is given by
\begin{equation} K(r) = 2 c_1^2 (D-1) D+\frac{c_2^2 (D-3) (D-2)^2 r^{2-2 D}}{D-1}~.\end{equation}
It is therefore clear that a black hole with the simplest radial function $W(r)=r$ cannot support a scalar field and cannot yield a regular black hole with non-trivial scalar field. One might also expect this behavior, since a simple self interacting scalar field theory does not satisfy $-\rho \neq p_r$ \cite{Jacobson:2007tj}.  A single-degree-of-freedom metric will result to $G_{t}^{~t} = G_{r}^{~r}$, which will further impose that $-\rho = p_r$, however this is not the case for a self interacting scalar field theory as we will discuss below.   A more complicated radial function will give rise to the scalar field and possibly a regular black hole.
Since we are interested in a regular black hole and not a wormhole spacetime, we should construct the geometry in a particular way in order to avoid a possible wormhole throat. To check for wormhole throats, we can compute the Kodama vector norm \cite{Faraoni:2021gdl,Chatzifotis:2021hpg}, which for our spacetime metric reads
\begin{equation} \label{wormcond} \partial_{\mu}g_{\theta\theta}\partial^{\mu}g_{\theta\theta}  =  g^{\mu\nu}\, \partial_{\mu}g_{\theta\theta}\, \partial_{\nu} g_{\theta\theta} = b(r)\, \left(\frac{dW(r)^2}{dr}\right)^2~.\end{equation}
A real positive root in the norm of the Kodama vector, corresponds to a black hole horizon, while a double root corresponds to a wormhole throat. It is clear that the double root can only be present in the $\left(dW(r)^2/dr\right)^2$ term. We are free to adjust a function and solve for the others, so taking a $W(r)$ that is always positive and monotonic, any root of the Kodama vector norm will always be sourced by $b(r)=0$, i.e a black hole horizon.
Under the light of the above discussion, we introduce
\begin{equation}
W(r) = \sqrt{r^2+A^2}~, \label{defor}
\end{equation}
where $A$ is a length scale, an ansatz which was at first considered in \cite{Bronnikov:2005gm}. By solving the remaining equations for a phantom scalar field
\begin{align}\label{phant}
f(\phi)=-1\,,
\end{align}
we can obtain all related functions as
\begin{equation}
\phi(r) = \sqrt{\frac{D-2}{\kappa}}\tan ^{-1}\left(\frac{r}{A}\right)~, \label{charge}
\end{equation}
\begin{multline}
b(r) = c_1 \left(A^2+r^2\right)-c_2 r \left(A^2+r^2\right)^{1-\frac{D}{2}} \left(\frac{r^2}{A^2}+1\right)^{D/2} \, _2F_1\left(\frac{1}{2},\frac{D}{2};\frac{3}{2};-\frac{r^2}{A^2}\right)
-2 (D-3) \left(A^2+r^2\right) \\
\left(\frac{r^2 \, _2F_1\left(\frac{1}{2},2-\frac{D}{2};\frac{3}{2};-\frac{r^2}{A^2}\right) \, _2F_1\left(\frac{1}{2},\frac{D}{2};\frac{3}{2};-\frac{r^2}{A^2}\right)}{A^4}-\int r \left(A^2+r^2\right)^{\frac{D-4}{2}-\frac{D}{2}} \left(\frac{r^2}{A^2}+1\right)^{D/2} \, _2F_1\left(\frac{1}{2},\frac{D}{2};\frac{3}{2};-\frac{r^2}{A^2}\right) \, dr\right)~,
\end{multline}
\begin{multline}
V(r) =\frac{(D-2) \left(A^2+r^2\right)^{-\frac{D}{2}-1} \left(\frac{r^2}{A^2}+1\right)^{-\frac{D}{2}}}{16 A^4 \pi }\times\\
\Bigg\{A^4 \left(\frac{r^2}{A^2}+1\right)^{D/2} \Bigg[\left(A^2+r^2\right)^{D/2} \left(-c_1 \left(A^2+(D-1) r^2\right)+D-3\right)+\frac{c_2 r \left(A^2+r^2\right) \left(A^2+(D-1) r^2\right) \, _2F_1\left(1,\frac{3-D}{2};\frac{3}{2};-\frac{r^2}{A^2}\right)}{A^2}\\
+c_2 r \left(A^2+r^2\right)\Bigg]+\frac{2 (D-3) r^2 \left(A^2+r^2\right)^{\frac{D}{2}+1} \left(\left(A^2+(D-1) r^2\right) \, _2F_1\left(1,\frac{3-D}{2};\frac{3}{2};-\frac{r^2}{A^2}\right)+A^2\right) \, _2F_1\left(\frac{1}{2},2-\frac{D}{2};\frac{3}{2};-\frac{r^2}{A^2}\right)}{A^2}\\
-2 A^4 (D-3) \left(A^2+r^2\right)^{D/2} \left(A^2+(D-1) r^2\right) \left(\frac{r^2}{A^2}+1\right)^{D/2} \left(\int \frac{ \, _2F_1\left(1,\frac{3-D}{2};\frac{3}{2};-\frac{r^2}{A^2}\right)r}{A^2 r^2+A^4} \, dr\right)
\Bigg\},
\end{multline}
where $_2 F_1(a,\,b\,;c\,;\,z)$ denote appropriate hypergeometric functions.
The above configurations solve the Einstein field equations along with the Klein-Gordon equation, which for a phantom scalar field and our metric ansatz yields
\begin{equation} b'(r) \phi '(r)+\frac{(D-2) b(r) W'(r) \phi '(r)}{W(r)}+b(r) \phi ''(r)+\frac{V'(r)}{\phi '(r)}=0~,\end{equation}
and can be obtained using equations (\ref{ttrr}), (\ref{ttuu}) and (\ref{v}).
The length scale $A$ has the role of a scalar charge since it controls the far-field behavior of the scalar field
\begin{equation} \label{scch} \phi(r\to\infty) \sim \sqrt{\frac{D-2}{8\pi}}\frac{\pi }{2 }-\sqrt{\frac{D-2}{8\pi}}\frac{A}{r}+\mathcal{O}\left(\left(\frac{1}{r}\right)^3\right)~.\end{equation}
We note that the presence of the scalar charge $A$ as it appears in relations  (\ref{charge}) and (\ref{scch}) also it appears in (\ref{defor}) indicating that matter deforms the geometry of the resulting black hole solution from the field equations.
The obtained spacetime is regular at the origin of the coordinate system. To see this, we expand asymptotically near the origin the metric functions to find
\begin{eqnarray}
&&W(r\to0) \sim A+ \frac{r^2}{2 A} + \mathcal{O}(r^4)~,\\
&&b(r\to0) \sim +A^2 c_1-c_2 r A^{2-D}+\frac{r^2 \left(A^2 c_1-D+3\right)}{A^2}+\frac{1}{6} c_2 (D-6) r^3 A^{-D} + \mathcal{O}(r^4)~.
\end{eqnarray}
Now using these expressions, the Kretschmann scalar is calculated near the origin and is found to be finite, however its expression is too complicated to be given here. The value for $r=0$ is simple thought and is given by
\begin{equation} K(r=0) = -\frac{8 c_1 (D-3)}{A^2}+\frac{2 (D-3) (3 D-8)}{A^4}+4 c_1^2 (D-1)~,\end{equation}
which is clearly finite. As a result one can conclude that the spacetime is regular.  We remind the reader, that the condition for regularity of the Kretschmann scalar is a necessary and sufficient condition for spacetime regularity \cite{Bronnikov:2022ofk}. Moreover, the matter content of the theory is also finite at $r=0$. Using the asymptotic expressions calculated before we find that the scalar field and potential yield
\begin{eqnarray}
&&\phi(r\to0) \sim \sqrt{\frac{D-2}{8\pi}}\frac{ r}{A }+\mathcal{O}\left(r^3\right)~,\\
&&V(r\to0) \sim -\frac{A^2 c_1 (D-2)-(D-3) (D-2)}{2 \left(8\pi A^2 \right)}+\frac{c_2 (D-2) r A^{-D}}{8\pi }+\mathcal{O}\left(r^2\right)~,
\end{eqnarray}
which are both finite at $r=0$. Hence the origin $r=0$ is point that belongs to our manifold. Now, having argued that the resulting spacetime is regular regardless of its dimensionality, we will discuss the behavior and nature of the matter threading the black hole, with the use of the energy-momentum tensor.
We can calculate the energy density, the radial pressure and the tangential pressure of the energy-momentum tensor as follows
\begin{eqnarray}
&&\rho = -T^{t}_{~t} = \frac{1}{2} b(r)f(\phi) \phi '(r)^2+V(\phi)~, \label{energy} \\
&&p_{r} = T^{r}_{~r} = \frac{1}{2} b(r)f(\phi) \phi '(r)^2-V(\phi)~, \label{radpre} \\
&&p_{\theta} = T^{\theta}_{~\theta} = -\frac{1}{2} b(r)f(\phi) \phi '(r)^2-V(\phi) = -\rho~. \label{tanpre}
\end{eqnarray}
Note here that $\rho \neq - p_r$. By combining these expressions we can discuss the energy conditions, namely the Weak Energy Condition (WEC) which states that the energy density seen by an observer with a timelike and future oriented $4-$velocity tangent vector $t^{\mu}$ is non-negative which implies that (\ref{energy}) should be non-negative i.e $\rho \ge 0$, the Null Energy Condition (NEC) which states that $\rho + p_{r}\ge0$ so that the geometry will have a focusing effect on null geodesics and the Strong Energy Condition (SEC) which implies that the geometry has a focusing effect on timelike geodesic congruences if $\rho+p_{i}\ge0$ for any $p_{i}$ and that $\rho + \sum_{i}p_{i} \ge0$. The nature of the scalar field theory dressing the black hole, can also be examined by calculating the equation of state
\begin{equation} w\equiv p_r/\rho~.
\end{equation}
The exact form of the solution of the field equations in terms of elementary functions for arbitrary dimension $D$ is unknown, hence we will discuss in detail the energy conditions for the special cases for various dimensions $D$ that we will discuss below. Here, we will point out some particular features that are present regardless of the dimensionality of spacetime. At the event horizon of the black hole, we have for the NEC and for the equation of state 
\begin{eqnarray}
&& \rho + p_r = -b(r_h) \phi '(r_h)^2 =0~,\\
&& w = \frac{p_r}{\rho} = \frac{-\frac{1}{2} b(r_h) \phi '(r_h)^2-V(r_h)}{-\frac{1}{2} b(r_h) \phi '(r_h)^2+V(r_h)} =- \frac{V(r_h)}{V(r_h)} = -1~.
\end{eqnarray}
Both of these equations are of particular interest. At first, the NEC holds at the event horizon of the black hole and after this it becomes negative driven by the phantom nature of the scalar field, since $b(r)>0$ for $r>r_h$. The equation of state at the event horizon of the black hole is exactly $-1$. This behavior is present even for a regular minimally coupled scalar field at the horizon. The consequence is that any hairy black hole that results from minimally coupled self-interacting scalar matter that has a finite scalar field at the horizon behaves in the same manner as a bare positive cosmological constant term at the event horizon of the black hole, a condition that is in favour of cosmological observations.

This is due the nature of the kinetic term $-\frac{1}{2} b(r_h) \phi '(r_h)^2$ at the horizon since $b(r_h)=0$ regardless of the dimensionality of spacetime. Hence, at the event horizon, a distant observer measuring the pressure to energy ratio cannot distinguish it from that of a positive cosmological constant. However, the radial pressure $p_r$ and energy density $\rho$ contain both the kinetic and potential energies of the scalar field which have no apparent reason to be given in terms of a cosmological constant at the horizon. The scalar field theory is finite and regular at the horizon.

\subsection{Violation of the  No-Hair Theorem } \label{hair}

A formal understanding of the existence of hair in our solutions comes by re-examining the way the standard no-hair theorem \cite{Bekenstein:1972ny} is violated in our case. To this, we commence our discussion by considering
the Klein-Gordon equation, which reads
\begin{equation}\Box\phi + \frac{dV}{d\phi}=0~.\end{equation}
Multiplying with $\phi$ and integrating over the exterior black hole region we have
\begin{equation}
\int d^Dx\sqrt{-g}\Big(\frac{1}{\sqrt{-g}}\partial_{\mu}(\phi\sqrt{-g}g^{\mu\nu}\partial_{\nu}\phi) - \nabla^{\mu}\phi\nabla_{\mu}\phi + \phi\frac{dV}{d\phi}\Big)=0~. \label{intKG}
\end{equation}

Now the first term of the above relation is usually considered to vanish in various hairy black holes, by assuming some fall-off behaviour for the scalar field at infinity. Let us now discuss the behavior of this total derivative term in our case, where we have a fixed scalar field fall-off for any dimension $D$. We can see that this term evaluates as
\begin{equation} \phi(r)\phi'(r)b(r)\sqrt{-g}\Big|_{r_h}^{\infty}~. \label{lim}\end{equation}

At the horizon we have that $b(r)=0$. At infinity, the metric can have an $b(r)\sim r^2$ asymptotic behavior i.e an asymptotically (A)dS behavior, since any term stronger than $r^2$ will make the curvature invariants diverge at infinity, rendering the solution ill-defined. Then it is easy to schematically show that the asymptotic behavior of the scalar field needed to cancel the contribution of the boundary term at infinity will be
\begin{equation} \phi(r\to\infty) \sim \mathcal{O}\left(\frac{1}{r^{(D+n-1)/2}}\right)~,\end{equation}
where $n$ is a real positive number. For a flat spacetime the condition will be
\begin{equation} \phi(r\to\infty) \sim \mathcal{O}\left(\frac{1}{r^{(D+n-3)/2}}\right)~.\end{equation}
Under this asymptotic behavior one can see that the contribution of (\ref{lim}) at infinity will also be null (when $n=0$ an effective cosmological constant term will emerge, which can be reabsorbed in the potential), and hence one can cancel the contributions from these total derivative terms. Our general result reduces for $D=5$ to the result of \cite{Farakos:2009fx}. In our case, the scalar field has a fixed asymptotic behavior regardless of the spacetime dimension, a situation that was not fixed ad hoc. Instead we  fixed the area function of the black hole. The scalar field has the desired asymptotic behaviour only for $D=3,4$ according to the No-Hair Theorem. Consequently, one cannot discard the total derivative term in our case for $D>4$. However it is not necessary to cancel the boundary term as long as the kinetic and potential terms of the theory are finite at large distances.  By discarding the boundary term, one can make some assumptions about the nature of the potential. It can be shown that for an asymptotically flat spacetime and for a regular scalar field, the potential has to be negative (at least for some region of $r$), in order to support the hairy structure \cite{potential}. By looking at (\ref{intKG}) one can deduce that for a phantom scalar field theory, a partly positive potential is needed in order to support a hairy structure (of course canceling the boundary term). In our solution for $D>4$, one cannot make any conclusions about the nature of the potential and as a result we have a violation of the No-Hair Theorem because of a relaxed asymptotic behavior in the scalar sector of our theory and not only because of the nature of the potential. This can also be true in a regular scalar field theory. In conclusion, for $D\ge 5$ the boundary term also contributes to the integral (\ref{intKG}), so the potential can be positive (negative) for a regular (phantom) scalar field and still violate the No-Hair theorem.

\section{Regular Black Hole Solutions in Various Spacetime Dimensions}\label{sec:cases}

In this Section we will explicitly discuss the regular black hole solution with scalar hair in various dimensions.  We will begin with $D=3$.

\subsection{The $D=3$ case}

Regular three dimensional black hole solutions were discussed in \cite{regular3}. For $D=3$, the solution of the field equations becomes
\begin{eqnarray}\label{d3sol}
&&b(r) = c_1 \left(A^2+r^2\right)-\frac{c_2 r \sqrt{A^2+r^2}}{A^2}~, \nonumber \\
&&\phi(r) = \frac{1}{\sqrt{8\pi }}\tan ^{-1}\left(\frac{r}{A}\right)~, \nonumber\\
&&V(r) = -\frac{2 A^2 c_1 r^2+A^4 c_1-2 c_2 r \sqrt{A^2+r^2}}{16\pi A^4 +16\pi A^2  r^2}~, \nonumber \\
&&f(\phi)=-1~,\nonumber\\
&&V(\phi) = \frac{c_2 \sin \left(\sqrt{8\pi } \phi \right)}{8\pi A^2  }+\frac{c_1 \cos \left(2 \sqrt{8\pi } \phi \right)}{32 \pi }-\frac{3 c_1}{32\pi }~.
\end{eqnarray}
The series expansion at infinity for the field $\phi(r)$, $r \to \infty$, reads
\begin{align}
\phi(r \to \infty) = \frac{\sqrt{\pi}}{4\sqrt{2} \, |A| } - \frac{1}{\sqrt{8\pi}}\, \frac{A}{r} + \mathcal O\left(\frac{A^3}{r^3}\right)\, ,
\end{align}
from which we conclude that $A$ (or, to be precise, $-\frac{A}{\sqrt{8\pi}}$) plays the r\^ole of a conserved scalar charge which was defined in
\eqref{scch} for a general $D$ dimensions.

For the metric function $b(r \to \infty)$, on the other hand, we have
\begin{equation} b(r\to\infty) \sim r^2 \left(c_1-\frac{c_2}{A^2}\right)+\left(A^2 c_1-\frac{c_2}{2}\right)+\frac{A^2 c_2}{8
   r^2}+\mathcal{O}\left(\left(\frac{1}{r}\right)^3\right)~,\end{equation}
which resembles the BTZ black hole \cite{BTZ} with corrections in the structure of spacetime that depend on the scalar charge $A$. The $\mathcal{O}(r^2)$ term is related to the cosmological constant $1/l^2$. Hence, we can identify
\begin{equation}\label{c1c2}
 c_1=\frac{c_2}{A^2}+\frac{1}{l^2}~.
 \end{equation}
Now to compute the mass we perform a coordinate transformation and we write the spacetime element as
\begin{eqnarray}
&&ds^2 = -B(R)dt^2 + \frac{R^2}{(R^2-A^2)B(R)}dR^2 +R^2d\theta^2~, \nonumber\\
&&B(R) = \left(\frac{c_2}{A^2}+\frac{1}{l^2}\right) R^2-\frac{c_2 R \sqrt{R^2-A^2}}{A^2}~. \label{gttgRR}
\end{eqnarray}
To compute the mass of the black hole we will use the quasilocal method \cite{quasilocal}. The quasilocal energy at finite $R$ reads
\begin{equation} e(R) = 2\left(\frac{1}{\sqrt{g^0_{RR}}}-\frac{1}{\sqrt{g_{RR}}}\right)~,\end{equation}
where $g^0_{RR}$ is a reference spacetime function which determines the zero of the energy. Now, the mass function at finite radius $R$ is given by
\begin{equation} M(R) = \sqrt{B(R)}e(R)~,\end{equation}
and at infinity, a distant observer measures a mass of
\begin{equation}\label{d3mass}
m = \frac{-2 A^2-c_2 l^2}{16 l^2}~.
\end{equation}
The scalar length scale $A$ enters the conserved mass and therefore it is a secondary hair of the black hole spacetime. Now, expanding the $g_{tt},g_{RR}$ functions asymptotically we find that
\begin{eqnarray}
&& g_{tt} \sim \frac{R^2}{l^2}-\frac{A^2+8 l^2 m}{l^2}-\frac{A^2 \left(A^2+8 l^2 m\right)}{4 l^2
   R^2}+\mathcal{O}\left(\left(\frac{1}{R}\right)^3\right)~, \nonumber\\
&& g_{RR}^{-1} \sim \frac{R^2}{l^2}-8 m+\frac{-40 A^2 l^2 m-A^4}{4 l^2 R^2}+\mathcal{O}\left(\left(\frac{1}{R}\right)^3\right)~.
\end{eqnarray}
The explicit asymptotic expressions for large distances, $ R \to \infty$, are given so one can see that the   mass and  the charge  appear in the solution.

The event horizon of the black hole is obtained by solving $b(r_h)=0$,
\begin{equation}\label{eventhor}
r_h = \frac{A^2+16 l^2 m}{\sqrt{3 A^2+32 l^2 m}}~,
\end{equation}
where we can see that $r_h>r_h^{\text{BTZ}}= 2\sqrt{2m}\,l$ since for small $A$ we have
\begin{equation} r_h(A\to0) \sim 2 \sqrt{2 \,m} \, l+\frac{A^2}{16 \sqrt{2}\, l \sqrt{m}}+\frac{3 A^4}{2048 \sqrt{2}\, l^3 m^{3/2}}+\mathcal{O}\left(A^6\right)~.\end{equation}
\begin{figure}[h]
\centering
\includegraphics[width=.40\textwidth]{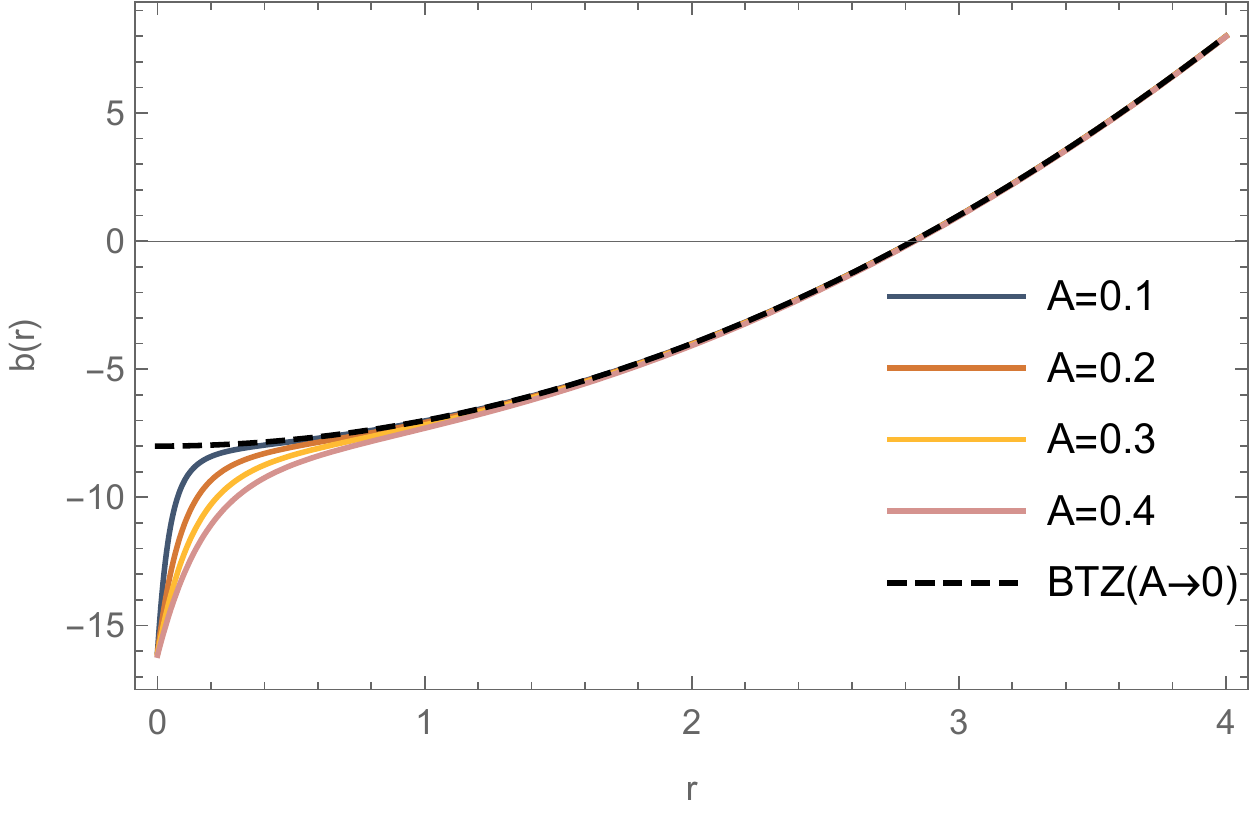}
\includegraphics[width=.40\textwidth]{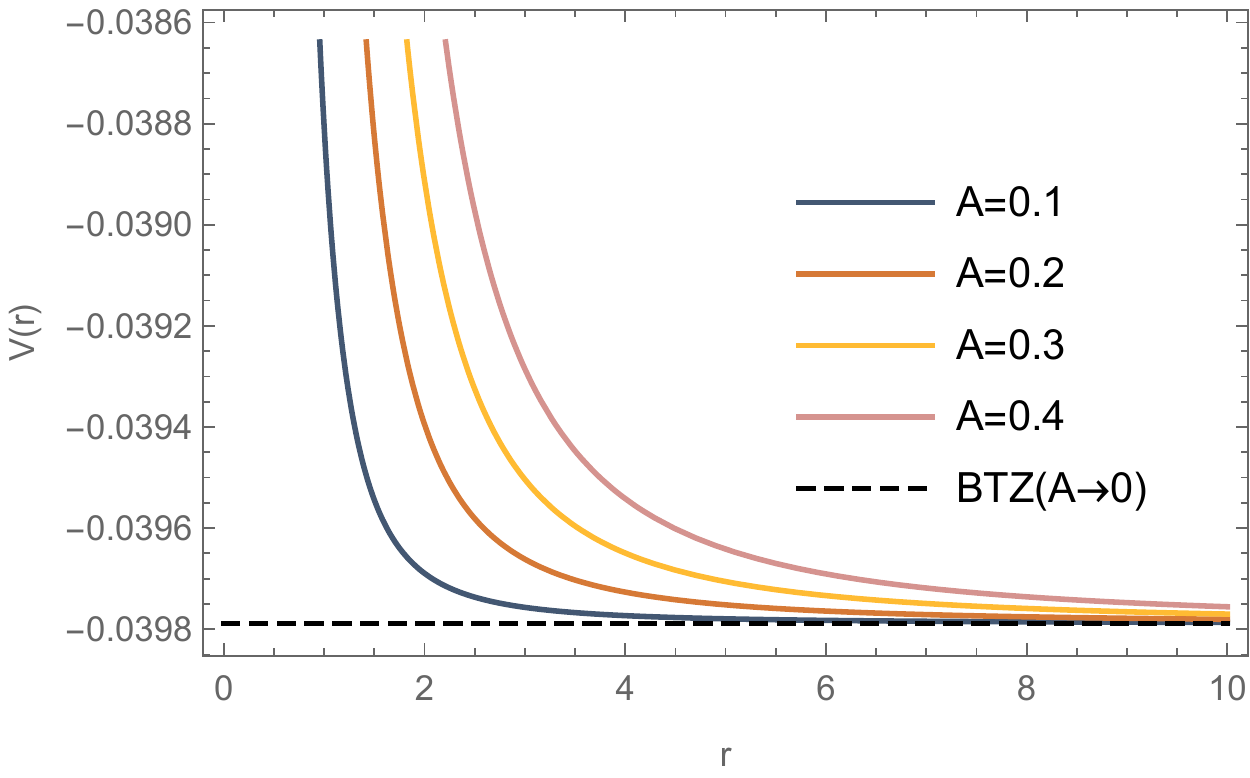}
\includegraphics[width=.40\textwidth]{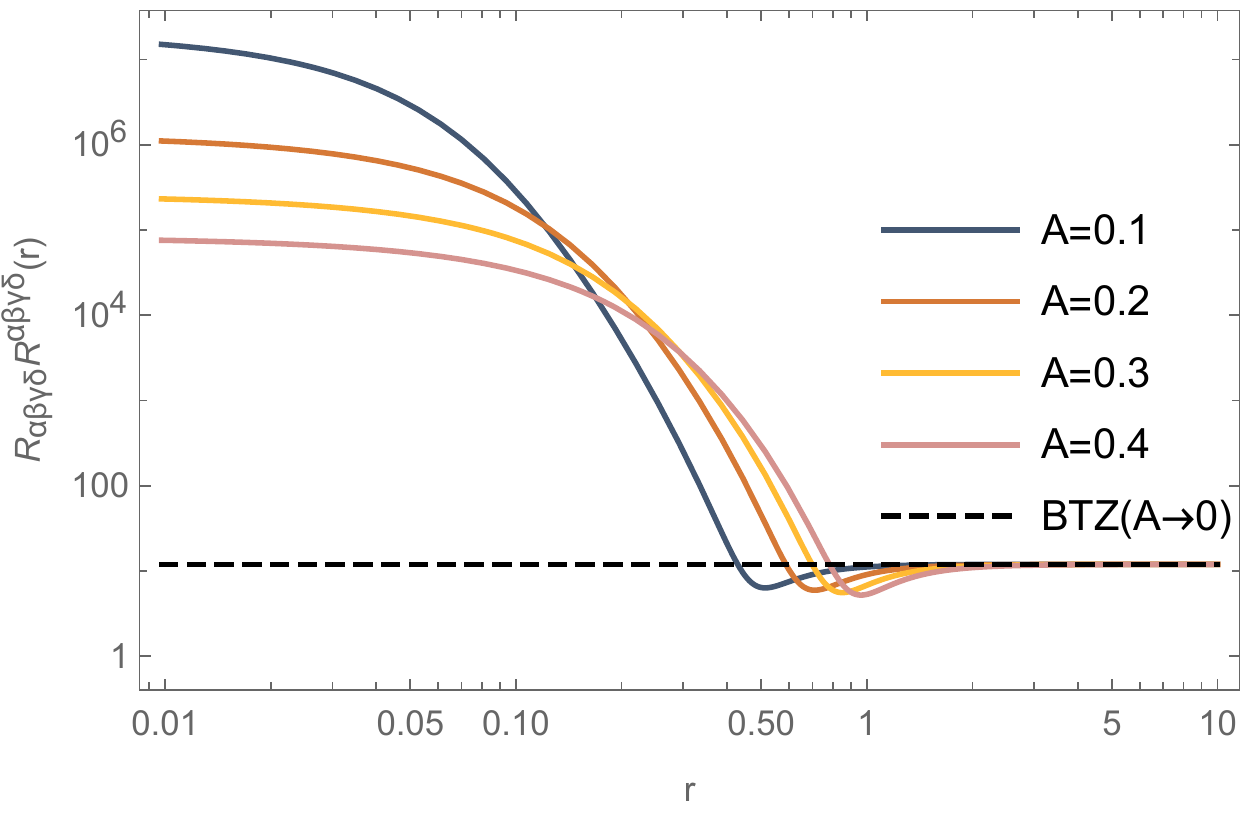}
\includegraphics[width=.40\textwidth]{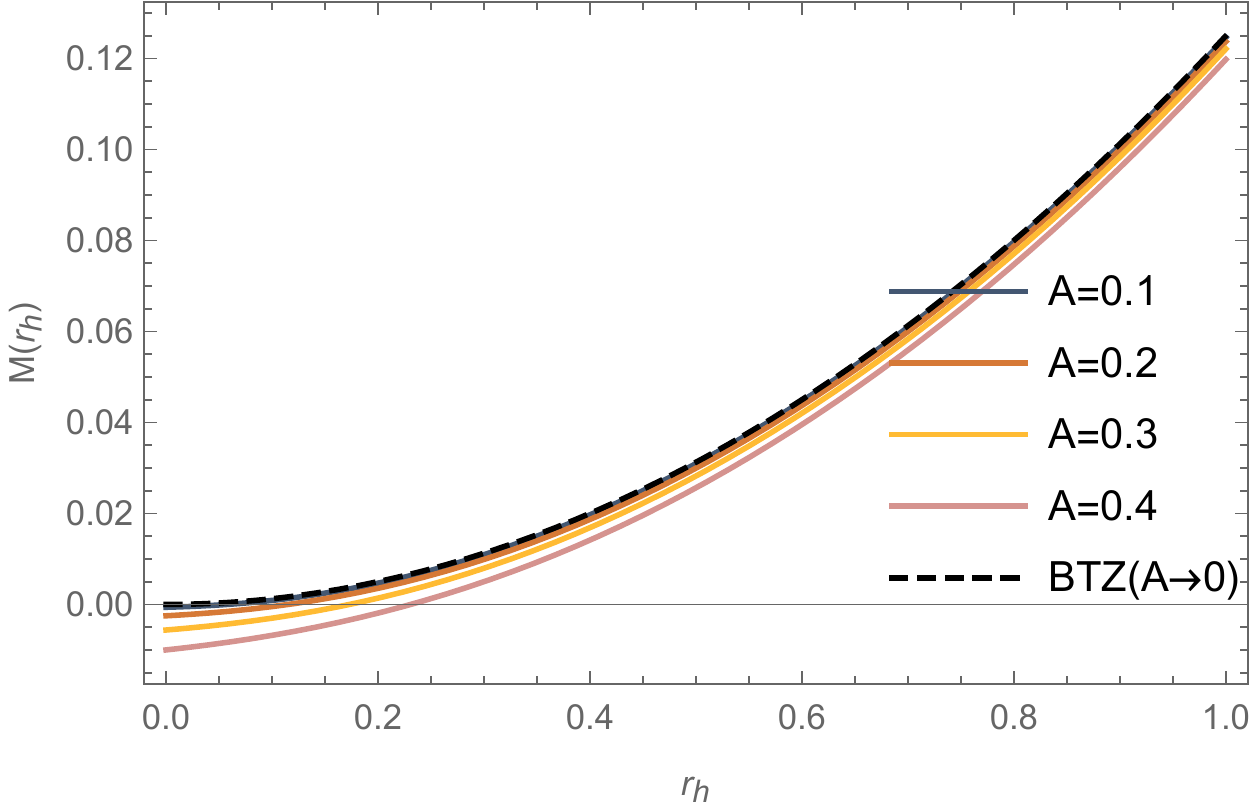}
\caption{The metric function $b(r)$, the potential $V(r)$, the square of the Riemann tensor and the mass as a function of the event horizon are ploted for $l=m=1$ while varying the scalar length scale $A$, alongside the BTZ black hole.}
\label{D3}
\end{figure}
In Fig.~\Ref{D3} we plot the metric function $b(r)$, the scalar potential $V(r)$ the square of the Riemann tensor $R_{\alpha\beta\gamma\delta}R^{\alpha\beta\gamma\delta}(r)$ and the mass function while varying the scalar charge $A$ along with the BTZ black hole case in order to compare. The BTZ metric and the hairy metric functions do not possess significant differences, which is expected. Both functions are regular at the origin,  the mass term is given by the angle deficit and is not a real mass term that will give rise to a non-zero Weyl tensor, a fact reflected also in the plot of the Riemann invariant.  The BTZ black hole is regular because there exists no pure gravitational term in the spacetime, since the graviton cannot propagate (not enough degrees of freedom). Hence the Riemann invariant is also regular at the origin of the BTZ black hole. In $D=3$ dimensions the only contributions to the Riemann tensor come from the Ricci tensor and the Ricci scalar which are trivially related to the AdS scale $l$ and hence in the BTZ case the Riemann square is constant. However, in the hairy cases, the Riemann tensor receives contributions from the phantom scalar field theory and is therefore dynamical. The origin $r=0$ is a regular point
\begin{align}\label{regular}
R_{\alpha\beta\gamma\delta}R^{\alpha\beta\gamma\delta}(r = 0)=8 \frac{\left(A^2+16 \,l^2 \,m\right)^2}{(A\, l)^4}\,.
\end{align}
It also possesses positive minima, which can be numerically evaluated, but nothing special happens at those minima, as far as the scalar field theory or the black hole spacetime are concerned. The mass function always increases with respect to the event horizon, while it does not develop any extremum, hence there will not be any first order phase transition.

We will now briefly discuss the thermodynamical properties of the solution. By performing a Wick rotation $t\to i\tau$ we can relate the periodicity of the Euclidean time $\beta$ with the inverse of the black hole temperature.  We will provide a calculation for the temperature via the Wick rotation. To do so, we will use the coordinate system (\ref{gttgRR}). We will at first ignore the angular part of the metric, and performing the rotation $t \to i\tau$, we're left with
\begin{equation} ds^2 = B(R)d\tau^2 +\frac{1}{F(R)}dR^2~,\end{equation}
where for notational simplicity we have set $1/F(R) = (R^2)/((R^2-A^2)B(R))$. Now, we expand the metric functions near the horizon
\begin{eqnarray}
&&B(R\to R_h) = B(R_h) + B'(R_h)(R-R_h) + ... = B'(R_h)(R-R_h)~,\\
&&F(R\to R_h) = F(R_h) + F'(R_h)(R-R_h) + ... = F'(R_h)(R-R_h)~,
\end{eqnarray}
and the reduced spacetime element reads
\begin{equation} ds^2 = B'(R_h)(R-R_h)d\tau^2 + \frac{1}{F'(R_h)(R-R_h)}dR^2~.\end{equation}
The above line element should cover the whole Euclidean space, and this will only be true if we treat $\tau$ as a periodic variable, elsewise, this metric will describe a cone. As a result we have to impose periodicity of $\tau$ as $\tau \sim \tau + 2\pi$, which comes from comparing our line element with the line element of polar coordinates $dS^2 = d\mathcal{R}^2 + \mathcal{R}^2d\Theta^2$, by identifying the following relations
\begin{eqnarray}
&&d\mathcal{R}^2 = \frac{1}{F'(R_h)(R-R_h)}dR^2 \to \mathcal{R} = 2\sqrt{\frac{R-R_h}{F'(R_h)}}~,\\
&&B'(R_h)(R-R_h)d\tau^2 =  \mathcal{R}^2d\Theta^2 \to \frac{\Theta}{\tau} = \frac{\sqrt{F'(R_h)B'(R_h)}}{2}~.
\end{eqnarray}
Now, $\Theta$ is periodic with period $2\pi$ and by denoting the period of $\tau$ with $\beta$ we find
\begin{equation} \frac{2\pi}{\beta} = \frac{\sqrt{F'(R_h)B'(R_h)}}{2} \to \frac{1}{\beta} \equiv T = \frac{\sqrt{F'(R_h)B'(R_h)}}{4\pi}~.\end{equation}
from which we can obtain the temperature as a function of the event horizon as
\begin{equation} \label{temp}
T(R_h) =\frac{A^2}{4 \pi  l^2 \left(R_h-\sqrt{R_h^2-A^2}\right)}~,\end{equation}
where we have substituted the mass parameter from the relation of the horizon $B(R_h)=0 \to m = M(R_h)$. It is clear that in the limit $A\to0$ one recovers the BTZ black hole temperature $T_{BTZ} = R_h/2 \pi  l^2$ \cite{BTZ}. A root in the derivative of the temperature will unveil the possibillity of a second order phase transition. To check this, we calculate the derivative in terms of $R_h$ which yields
\begin{equation} \frac{dT(R_h)}{dR_h}=\frac{A^2}{\sqrt{R_h^2-A^2} \left(4 \pi  l^2 R_h-4 \pi  l^2 \sqrt{R_h^2-A^2}\right)}~,\end{equation}
where no root is present for $R_h>A$. The mass function reads
\begin{equation} M(R_h) = \frac{R_h \sqrt{R_h^2-A^2}-2 A^2+R_h^2}{16 l^2} \to \frac{dM(R_h)}{dR_h}= \frac{\left(\sqrt{R_h^2-A^2}+R_h\right)^2}{16 l^2 \sqrt{R_h^2-A^2}}~,\end{equation}
which is clearly always positive.

The entropy may be obtained by using Wald's formula \cite{Wald:1993nt} which reads
\begin{equation} \mathcal{S} = -2\pi \int _{\Sigma_1} \frac{\partial \mathcal{L}}{\partial R_{\alpha\beta\gamma\delta}} \hat{\varepsilon}_{\alpha\beta}\hat{\varepsilon}_{\gamma\delta}~,\end{equation}
where $\mathcal{L}$ denotes the Lagrangian of our theory, $\hat{\varepsilon}_{\alpha\beta}$ is the binormal vector and the integral is evaluated on the horizon circumference. Performing the integration we find (restoring the units of the gravitational constant $G$ for completeness and clarity)
\begin{equation} \mathcal{S} = \frac{4\pi^2 W(r_h)}{\kappa} = \frac{\mathcal{A}}{4G}~,\end{equation}
where $\mathcal{A}=2\pi W(r_h)=2\pi R_h$ is the circumference of the black hole.
The heat capacity can now be evaluated as
\begin{equation} C(R_h) = T\frac{dS}{dT} = \frac{1}{2} \pi  \sqrt{R_h^2-A^2}~,\end{equation}
where no divergent point is present so the hairy black holes are thermally stable.
%\color{black}

Now, regarding the phantom matter threading the black hole spacetime, we will discuss the energy conditions. The explicit expressions for the energy density and radial pressure, and their sum read respectively,
\begin{eqnarray}
&& \rho(r) = \frac{A^2 \left(16 l^2 m\sqrt{A^2+r^2}+r^2 \sqrt{A^2+r^2}-24 l^2 r m-2 r^3\right)+16 l^2 r^2 m \left(\sqrt{A^2+r^2}-r\right)+A^4 \left(\sqrt{A^2+r^2}-3 r\right)}{8\pi A^2  l^2 \left(A^2+r^2\right)^{3/2}}~, \nonumber\\
&&p_r(r) = \frac{r \left(A^2 \left(-r \sqrt{A^2+r^2}+8 l^2 m+2 r^2\right)+16 l^2 r m \left(r-\sqrt{A^2+r^2}\right)+A^4\right)}{8\pi A^2   l^2 \left(A^2+r^2\right)^{3/2}}~, \nonumber \\
&&\rho(r) + p_r(r) = \frac{16 l^2 m \left(\sqrt{A^2+r^2}-r\right)+A^2 \left(\sqrt{A^2+r^2}-2 r\right)}{8 \pi  l^2 \left(A^2+r^2\right)^{3/2}}~.
\end{eqnarray}
We can also calculate the equation of state $w=p_r/\rho$ obtaining
\begin{equation} w(r) = -\frac{r \left(A^2 \left(-r \sqrt{A^2+r^2}+8 l^2 m+2 r^2\right)+16 l^2 r m \left(r-\sqrt{A^2+r^2}\right)+A^4\right)}{A^2 \left(-16 l^2 m \sqrt{A^2+r^2}-r^2 \sqrt{A^2+r^2}+24 l^2 r m+2 r^3\right)+16 l^2 r^2 m \left(r-\sqrt{A^2+r^2}\right)-A^4 \left(\sqrt{A^2+r^2}-3 r\right)}~, \end{equation}
with the asymptotic expressions near the origin and at large distances yielding
\begin{eqnarray}
&&w(r\to0)\sim\frac{r \left(A^2+8 l^2 m\right)}{A \left(A^2+16 l^2 m\right)}+\frac{2 r^2 \left(8 A^2 l^2 m+A^4-32 l^4
   m^2\right)}{\left(A^3+16 A l^2 m\right)^2}+\mathcal{O}\left(r^3\right)~, \nonumber \\
&&w(r\to\infty) \sim -1+\frac{A^2}{r^2}+\mathcal{O}\left(\left(\frac{1}{r}\right)^4\right)~.
\end{eqnarray}
To understand better the nature of matter threading the black hole we will depict the above quantities.
\begin{figure}[h]
\centering
\includegraphics[width=.40\textwidth]{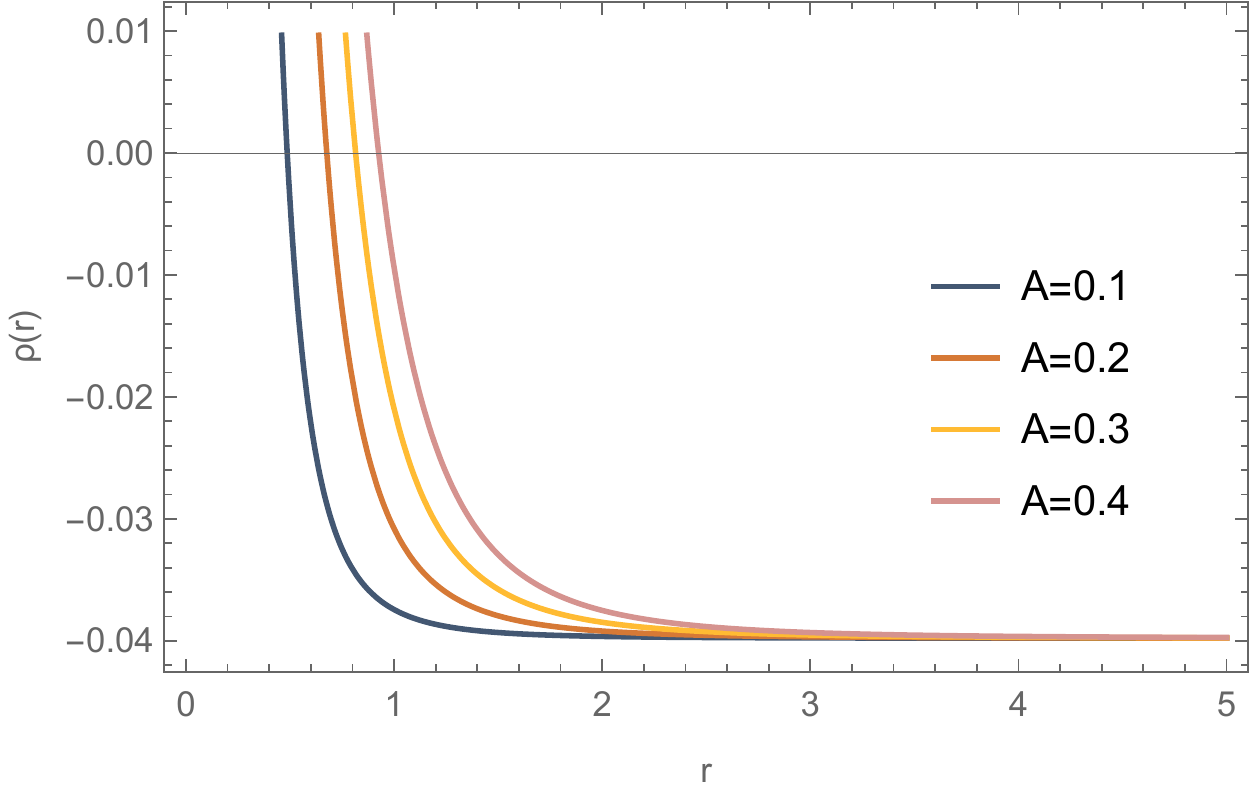}
\includegraphics[width=.40\textwidth]{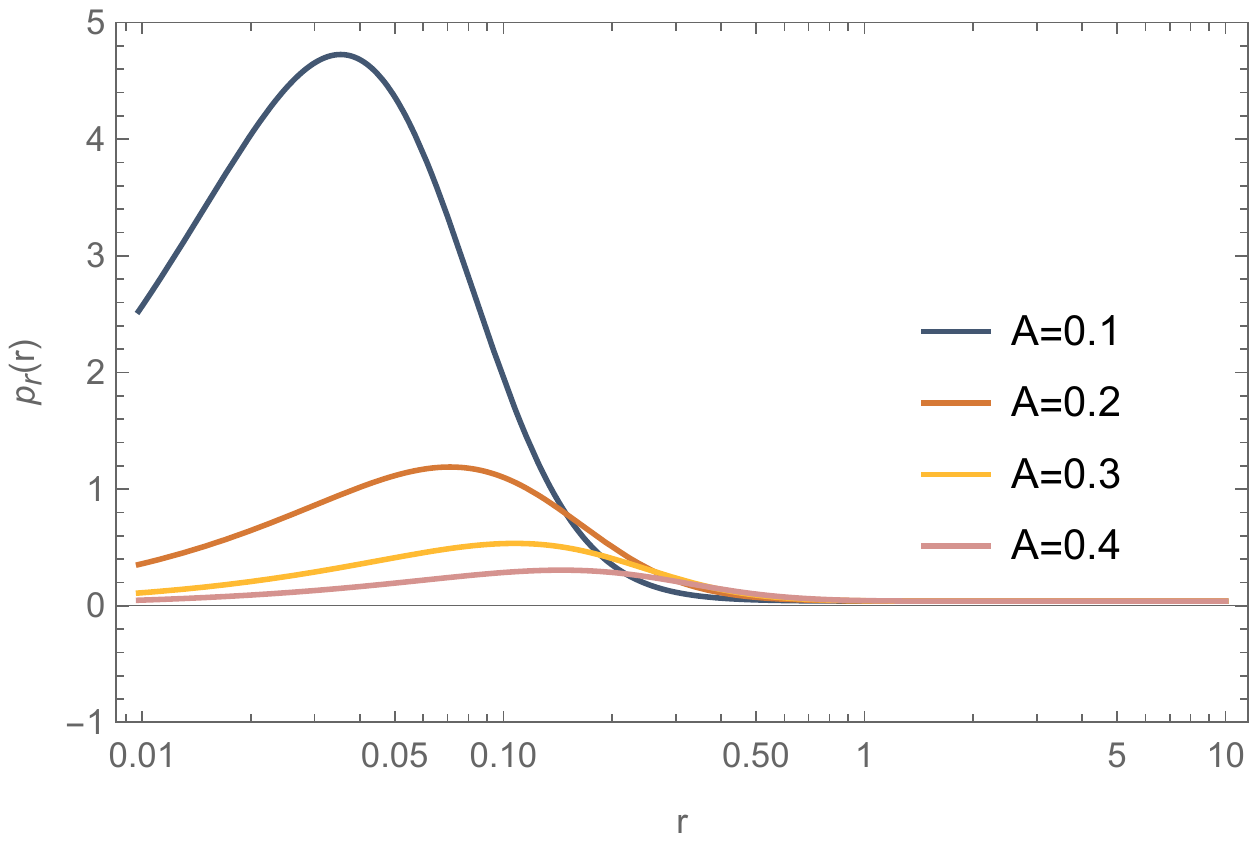}
\includegraphics[width=.40\textwidth]{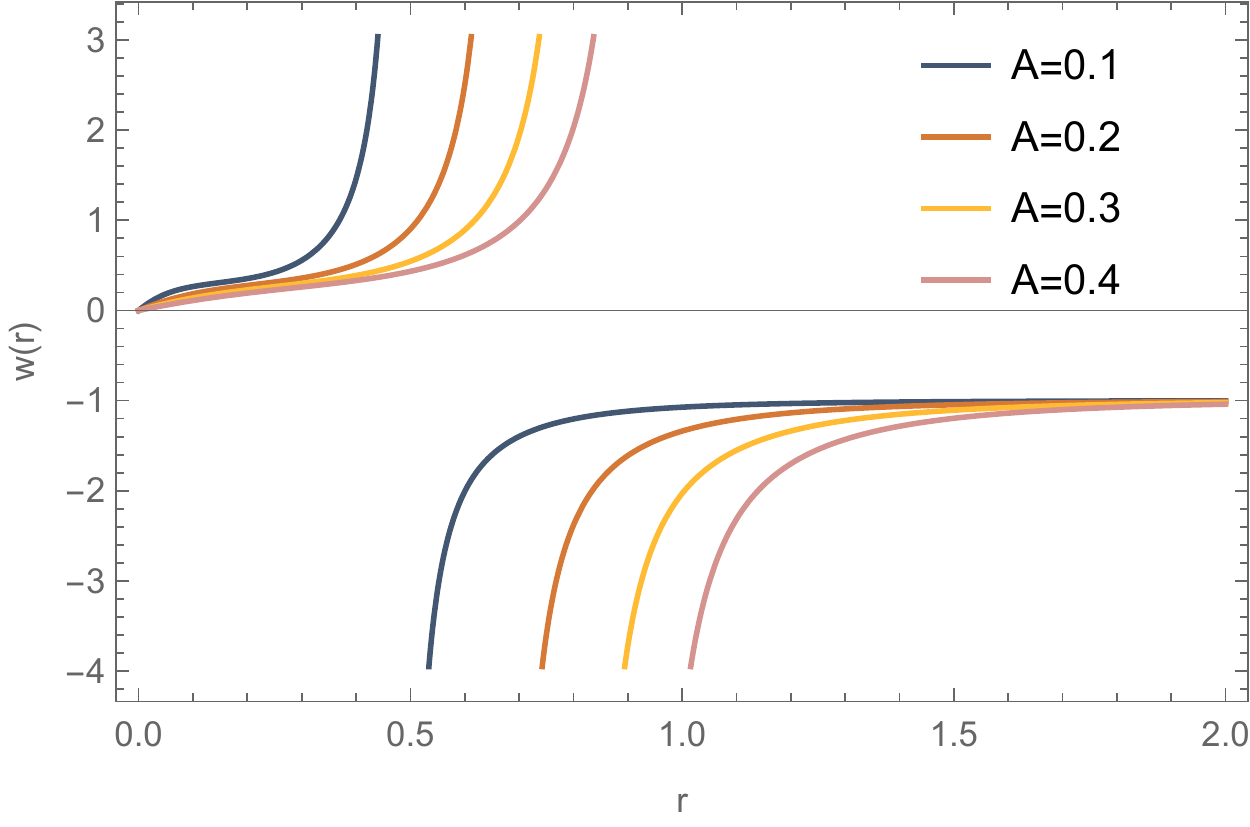}
\includegraphics[width=.40\textwidth]{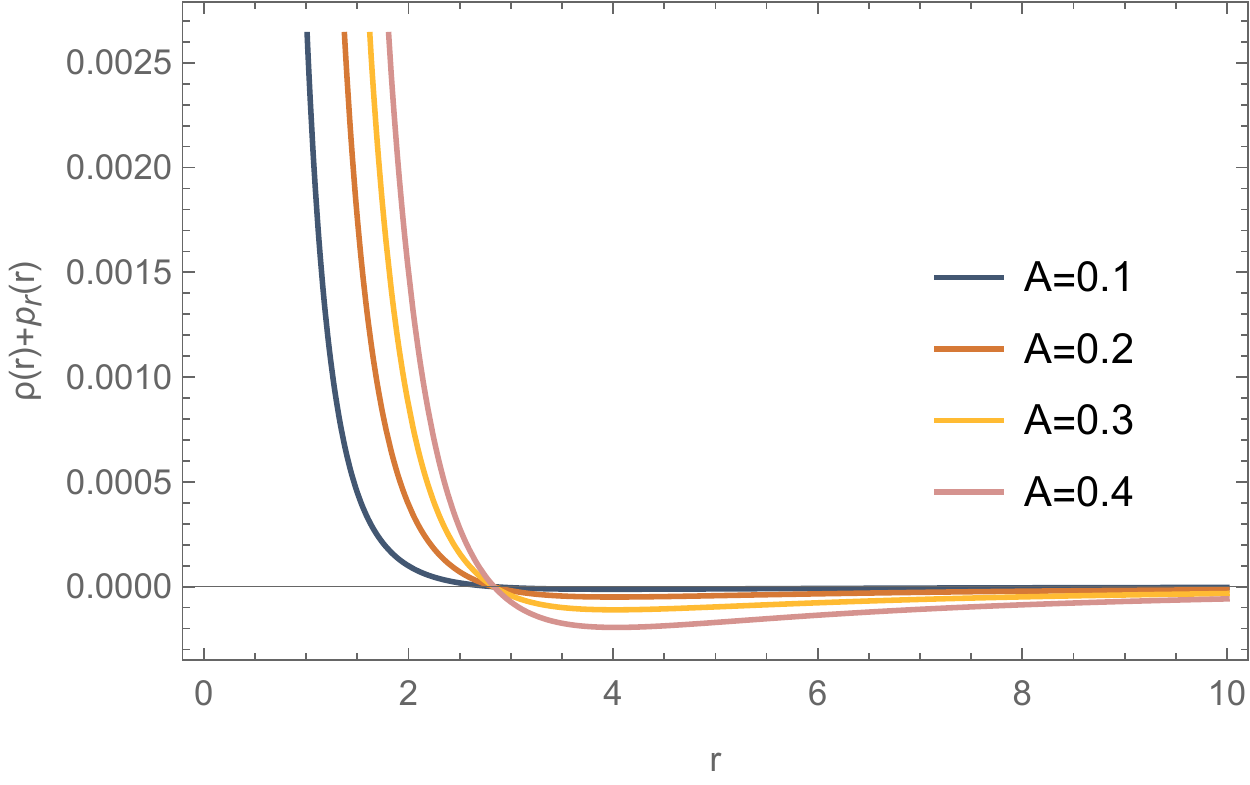}
\caption{The energy density $\rho(r)$, the radial pressure $p_r(r)$, the equation of state $w(r)$ and the sum of the energy density and pressure $\rho(r)+p_r(r)$ for $m=l=1$, while varying the scalar length scale $A$ for the three dimensional case.}
\label{matterD3}
\end{figure}
From Fig. \ref{matterD3} we can see that $\rho(r)$ changes sign at a point inside the black hole event horizon, signalising in this way a divergent point in the energy density, however nothing special happens at this point, except from the fact that the matter content of the theory becomes null there. It is also clear that the WEC is violated at the black hole horizon and in the causal region of spacetime, due to the phantom nature of the matter field. The radial pressure develops a global maximum and a local minimum which can be obtained numerically and the equation of state becomes divergent at the point where the matter content of the theory becomes null, while at larges distances asymptotes to a pure de Sitter case. The NEC is satisfied inside the black hole and at the event horizon and violated for any $r>r_h$ due to the phantom nature of the scalar field.

 In conclusion, we have obtained an analog of the BTZ black hole dressed with a secondary, phantom scalar hair. The phantom scalar field theory has a well defined behavior near the origin, by introducing a length scale which removes the central singularity. However, when the scalar field is regular one can see that ill-defined behaviors near the origin sourced by the scalar field function or the scalar potential will result to a black hole with a singularity \cite{singularbtz}.

As we already discussed the resent astrophysical observations indicate a possible connection of the singularity-free  BH with the cosmological evolution. The early cosmological evolution is governed by dark energy which is generated by a phantom scalar field which connect the mass of the BH with dark energy, defining in this way the black Universe. In our study the presence of phantom matter deforms the geometry and as it is showed  in relation \eqref{d3mass} it is connected with the mass of the black hole. Motivated by this we will impose a relation between the black hole mass and the scalar charge of the form \begin{align}\label{cond}
\frac{m}{A^2} = |\xi | \, G \,,
\end{align}
where $\xi \in \mathbb R$ is an arbitrary real dimensionless constant parameter, and we reinstated the units of the gravitational constant. In  three dimensions the mass is a dimensionless quantity,  and  the fundamental constant
of the BTZ solution~\cite{BTZ} is the AdS scale, and this gives rise to the event horizon not the mass, so that $m/A^2 \sim [\rm length]^{-2}$.
This relation guarantees that the potential is independent of the BH mass and scalar charge. Using relations \eqref{c1c2} and
\eqref{d3mass}, we may write $V(\phi)$ as
\begin{align}\label{d3pot}
V(\phi) &= \frac{m}{A^2} \Big[ \frac{2}{\pi}\, {\rm sin}(\sqrt{8\pi}\, \phi) + \frac{1}{2\pi} \, {\rm cos}(2\sqrt{8\pi}\, \phi) - \frac{3}{2\pi}\Big] + \frac{1}{4\pi\, l^2} {\rm sin}(\sqrt{8\pi}\, \phi) + \frac{3}{32\pi\, l^2} \, {\rm cos}(2\sqrt{8\pi}\, \phi) - \frac{9}{32\pi\, l^2}\,, \\
 &= |\xi | \, G \Big[ \frac{2}{\pi}\, {\rm sin}(\sqrt{8\pi}\, \phi) + \frac{1}{2\pi} \, {\rm cos}(2\sqrt{8\pi}\, \phi) - \frac{3}{2\pi}\Big] + \frac{1}{4\pi\, l^2} {\rm sin}(\sqrt{8\pi}\, \phi) + \frac{3}{32\pi\, l^2} \, {\rm cos}(2\sqrt{8\pi}\, \phi) - \frac{9}{32\pi\, l^2}\,.
\end{align}
The condition \eqref{cond} implies that the hair is secondary but,
an important question arises, regarding the effect of this on the thermodynamics of the respective compact object.
To this end, we remark that in three dimensional General Relativity, the thermodynamics of the BTZ black hole is entirely different from that of the (3+1)-dimensional case. First,
 temperature of the BTZ black hole is given by
$$T(m) \sim m^{1/2}~,$$
so the temperature decreases as the black hole evaporates.\footnote{The reader should take notice of the fundamental difference from the four dimensional case, where $T(m) \sim m^{-1}\,.$} Moreover, Stefan's law of radiation is schematically expressed in this case by
$$\frac{dm}{dt} \sim T^4 \sim m^2 \to t \sim m^{-1}~,$$
hence it seems that the BTZ black hole will take an infinite amount of time to evaporate, if the black hole does indeed evaporate \cite{Reznik:1994py}. Therefore,  we can define the ratio \eqref{cond} as a new parameter of the theory giving a fixed mass to scalar charge ratio, so that the scalar potential is independent of the mass and charge of the compact object. So when $m$ is changing and the black hole radiates, the scalar charge also changes in order to keep the ratio $\xi$ constant. This is also consistent with the black Universe, in which a change in the dark energy implies also a change in the mass of the BH.

Concluding the discussion on the $D=3$ case, we will briefly discuss the first law of thermodynamics. As we have already discussed in the introduction, inconsistencies appear when one identifies the black hole mass as the internal energy of the black hole. This behavior is rooted in the fact that the matter Lagrangian is not entirely independent of the black hole mass. One can easily check that the first law of thermodynamics expressed as $dM = TdS$ does not hold in our solution. Since the entropy of the black hole obtained from the Euclidean path integral and Wald's formula should coincide (in our case both these methods give the Bekenstein-Hawking area law) and the temperature is a kinematic effect, independent of the theory under consideration, we came to the conclusion that one has to consider that the internal energy of the black hole is not given by the conserved black hole mass, hence $m \neq E$. The parameter of our theory is $\xi$. So we can replace $m$ with $\xi$ through the relation (\ref{cond}), and now the horizon radius can be obtained as
\begin{equation} r_h = \frac{A \left(16 l^2 \xi +1\right)}{\sqrt{32 l^2 \xi +3}}~,
\end{equation}
where it is clear that in this expression the only parameter that can vary is $A$. Therefore we have
\begin{equation} T(r_h)dS(r_h) = T(r_h)\frac{\partial S}{\partial r_h}dr_h = T(r_h)\frac{\partial S}{\partial r_h}\frac{\partial r_h}{\partial A}dA = T(r_h)\frac{\partial S}{\partial A}dA = \frac{A \left(16 l^2 \xi +1\right)}{8 l^2}dA~,
\end{equation}
and now through the first law $dE=TdS$ we can obtain the internal energy of the black hole at the horizon as
\begin{equation} dE = \frac{A \left(16 l^2 \xi +1\right)}{8 l^2}dA \to E(A) = \frac{A^2}{16 l^2}+A^2 \xi = m +\frac{A^2}{16 l^2}~,\end{equation}
where we can see that $E \neq m$ while when $A\to0$ we have that $E=m$ which is the case of the BTZ black hole.

\subsection{The $D=4$ case}

In this subsection we carry out an analytical study of the  $D=4$ regular black hole solution first discussed  in \cite{Bronnikov:2005gm}, calculating also the energy conditions. The stability of the system against radial perturbations was also discussed \cite{Bronnikov:2012ch}, where it was found that the solutions are stable for particular values of the constants. Some properties of this solution like the gravitational lensing and the accretion process can be found in \cite{Ditta:2020jud, Ding:2013vta}.  The solution reads
\begin{eqnarray}\label{bhor}
&&b(r) = c_1 \left(A^2+r^2\right)-\frac{c_2 \left(\left(A^2+r^2\right) \tan ^{-1}\left(\frac{r}{A}\right)+A r\right)+2 A r^2}{2 A^3}~,
\end{eqnarray}
and
\begin{eqnarray}\label{d4solution}
&&\phi(r) = \frac{1}{2 \sqrt{\pi }}\tan ^{-1}\left(\frac{r}{A}\right)~, \nonumber \\
&&V(r) = \frac{c_2 \left(\left(A^2+3 r^2\right) \tan ^{-1}\left(\frac{r}{A}\right)+3 A r\right)-2 A \left(A^2 c_1-1\right) \left(A^2+3 r^2\right)}{16 \pi  A^3 \left(A^2+r^2\right)}~, \nonumber \\
&&V(\phi) = \frac{4 A \left(A^2 c_1-1\right) \left(\cos \left(4 \sqrt{\pi } \phi \right)-2\right)+c_2 \left(3 \sin \left(4 \sqrt{\pi } \phi \right)-4 \sqrt{\pi } \phi  \left(\cos \left(4 \sqrt{\pi } \phi \right)-2\right)\right)}{32 \pi  A^3}~.
\end{eqnarray}
The series expansion for the scalar field at infinity, again leads to
\begin{align}
\phi(r \to \infty) = \frac{\sqrt{\pi}}{4\sqrt{2} \, |A| } - \frac{1}{2\sqrt{\pi}}\, \frac{A}{r} + \mathcal O\left(\frac{A^3}{r^3}\right)\, ,
\end{align}
from which we conclude that $A$ (or, to be precise, $-\frac{A}{2\sqrt{\pi}}$) plays the r\^ole of a conserved scalar charge.

The metric function $b(r)$ at infinity reads
\begin{equation} b(r\to\infty) \sim r^2 \left(c_1-\frac{4 A+\pi  c_2}{4 A^3}\right)+\left(A^2 c_1-\frac{\pi  c_2}{4 A}\right)+\frac{c_2}{3 r}-\frac{A^2 c_2}{15
   r^3}+\mathcal{O}\left(\left(\frac{1}{r}\right)^5\right)~,\end{equation}
which resembles the (A)dS Schwarzschild black hole with corrections in the structure of spacetime that depend on the scalar charge $A$. By applying a transformation of the form $r^2 = R(r)^2 -A^2$, where $R$ will be the new coordinate we find that
\begin{equation} b(R\to\infty) \sim R^2 \left(-\frac{\pi  c_2}{4 A^3}-\frac{1}{A^2}+c_1\right)+1+\frac{c_2}{3 R}+\mathcal{O}\left(\left(\frac{1}{R}\right)^3\right)~.\end{equation}
It is clear that there is no deficit angle at large distances and the solution describes a pure singularity-free black hole and not a gravitational monopole.
To make the spacetime asymtpotically flat, we may set
\begin{equation} A^2 c_1-\frac{\pi  c_2}{4 A}=1 \to c_1=\frac{4 A+\pi  c_2}{4 A^3}~,\end{equation}
which fixes the value of $c_1$. Now the asymptotic relation yields
\begin{equation} b(r) \sim 1+\frac{c_2}{3 r}-\frac{A^2 c_2}{15 r^3}+\frac{A^4 c_2}{35 r^5}+\mathcal{O}\left(\left(\frac{1}{r}\right)^7\right)~.\end{equation}
The metric above is clearly asymptotically flat, and resembles the Schwarzschild black hole, while corrections in the structure of spacetime appear as $\mathcal{O}(r^{-n})$ terms (where $n \ge 3$) and are sourced by the (conserved) phantom scalar charge $A$. The other conserved charge  is the black hole mass, which can be calculated using the Komar integral. To compute the mass term we rewrite the line element in the form
\begin{eqnarray}
&&ds^2 = -B(R)dt^2 +B(R)^{-1}\left(\frac{R^2}{R^2-A^2}\right)dR^2 + R^2d\Omega^2~, \nonumber \\
&&B(R) =1- \frac{ c_2 \sqrt{R^2-A^2}}{3A^2}+\frac{ c_2 R^2}{6 A^3} \left(\pi -2 \cot ^{-1}\left(\frac{A}{\sqrt{R^2-A^2}}\right)\right)~.
\end{eqnarray}
The Komar mass $m$ is given by
\begin{equation} m = -\frac{1}{8\pi} \lim_{R\to\infty} \oint_{\partial \Sigma} \nabla^{\mu}\zeta^{\nu}dS_{\mu\nu} = \frac{1}{4\pi}\lim_{R\to\infty}\oint_{\partial \Sigma}dA n_{\mu}u_{\nu}\nabla^{\mu}\zeta^{\nu}~, \label{komar} \end{equation}
where $\zeta^{\mu}$ is the Killing vector field associated with time translation and energy conservation, since the spacetime element is independent of $t$, $\zeta^{\mu} = (1,0,0,0)$, $n^{\mu} = (B(R)^{-1/2},0,0,0)$ is the future oriented timelike unit vector and $u^{\mu}=(0,F(R)^{1/2},0,0)$ is the unit normal on the boundary, which is taken to be a two sphere with an infinite radius and $1/F(R) = (R^2)/((R^2-A^2)B(R))$. The mass can then be computed as
\begin{equation}\label{d4mass}
m = \frac{1}{4\pi} \lim_{R\to\infty} \oint_{\partial \Sigma}\left(dA n_{\mu}u_{\nu}g^{\mu\kappa}\Gamma^{\nu}_{\kappa\sigma}\zeta^{\sigma}\right)dA =\lim_{R\to\infty} \frac{1}{2} R^2 \sqrt{1-\frac{A^2}{R^2}} B'(R) = -\frac{c_2}{6}~,
\end{equation}
and is clearly not affected by the scalar field. Then,  if one treats $m$ as independent of the scalar charge, the obtained spacetime describes a regular, asymptotically flat black hole, with a primary phantom scalar hair.
Compared with the previously described  $D=3$ example, \eqref{d3mass}, this would imply that the nature of the scalar hair depends on the dimensionality of spacetime, with the hair being secondary for odd and primary for even $D$ (indeed, this can be confirmed by the explicit calculations for $D \ge 5$ performed in  the next section, if one treats mass and scalar charge as independent.)

\begin{figure}[h]
\centering
 \includegraphics[width=.40\textwidth]{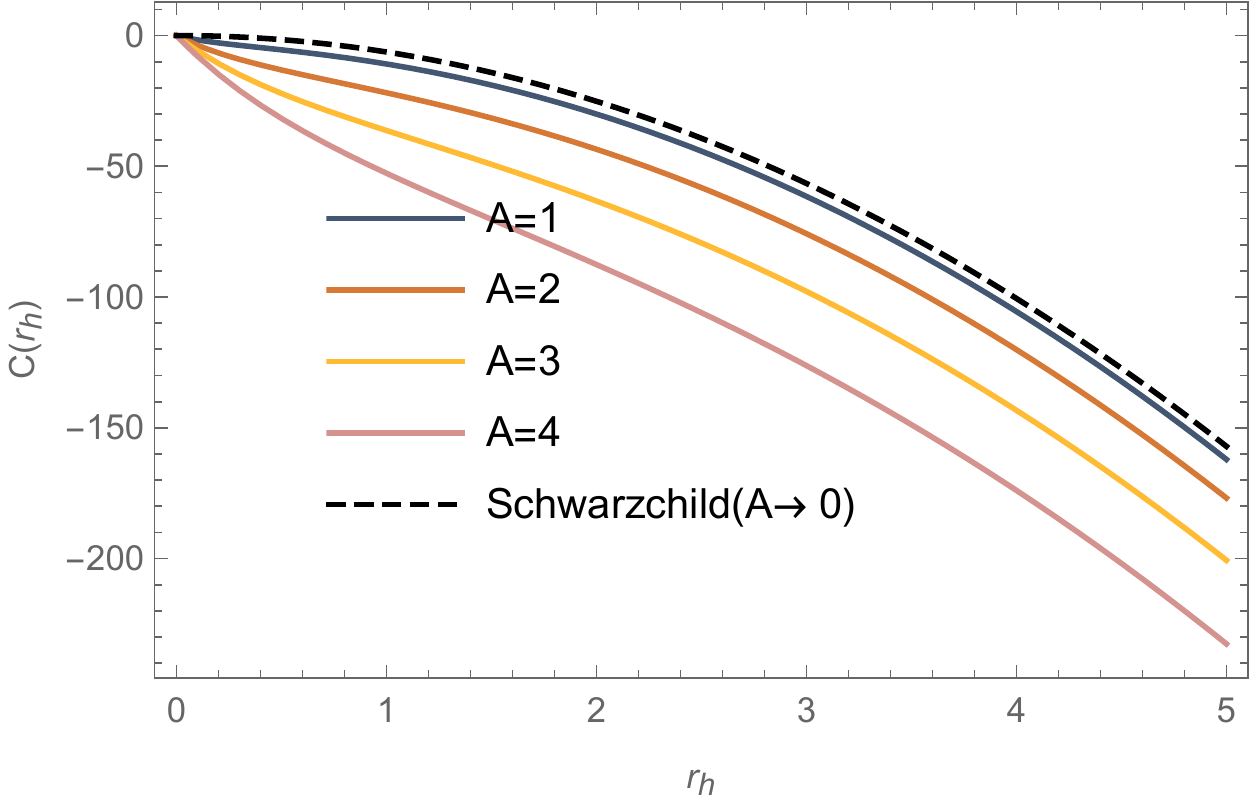}
\caption{The heat capacity $C(r_h) = TdS/dT\bigg|_{R_h\to\sqrt{r_h^2+A^2}}$ as a function of the horizon radius.} \label{heat4}
\end{figure}

\begin{figure}[h]
\centering
\includegraphics[width=.40\textwidth]{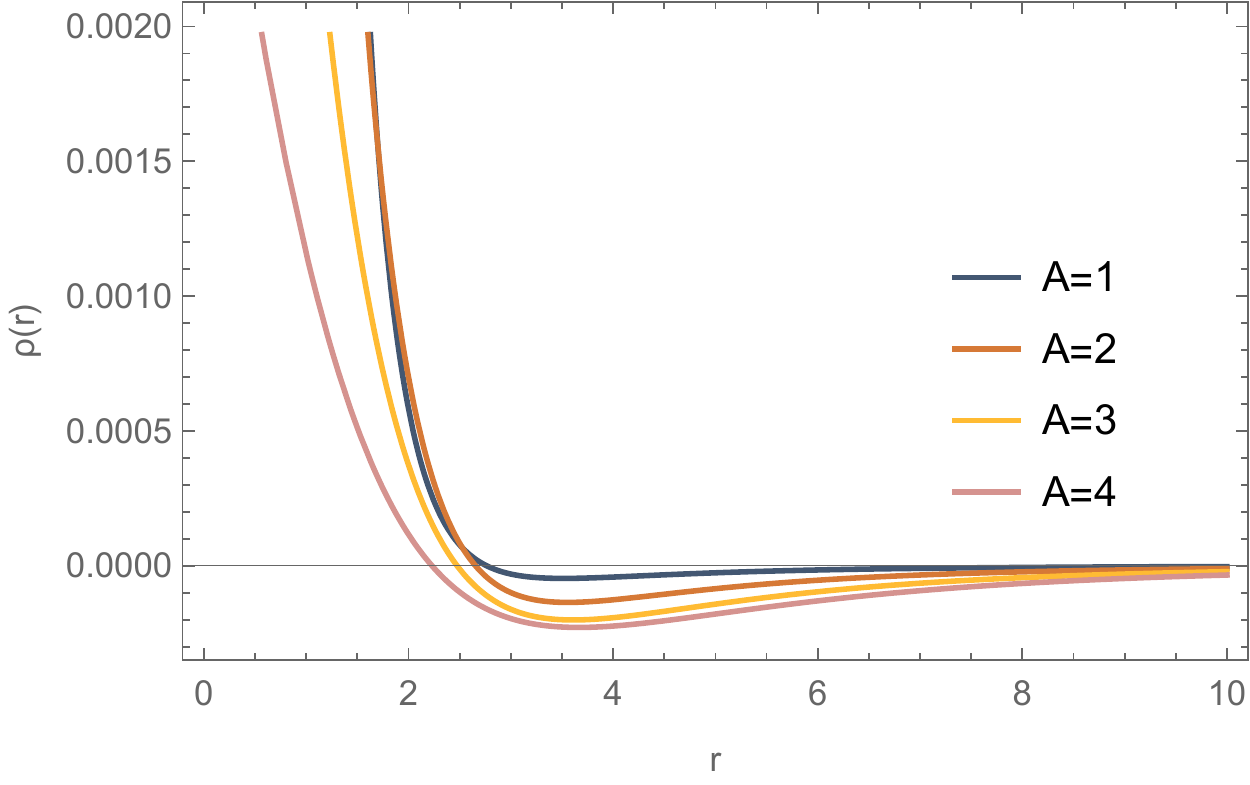}
\includegraphics[width=.40\textwidth]{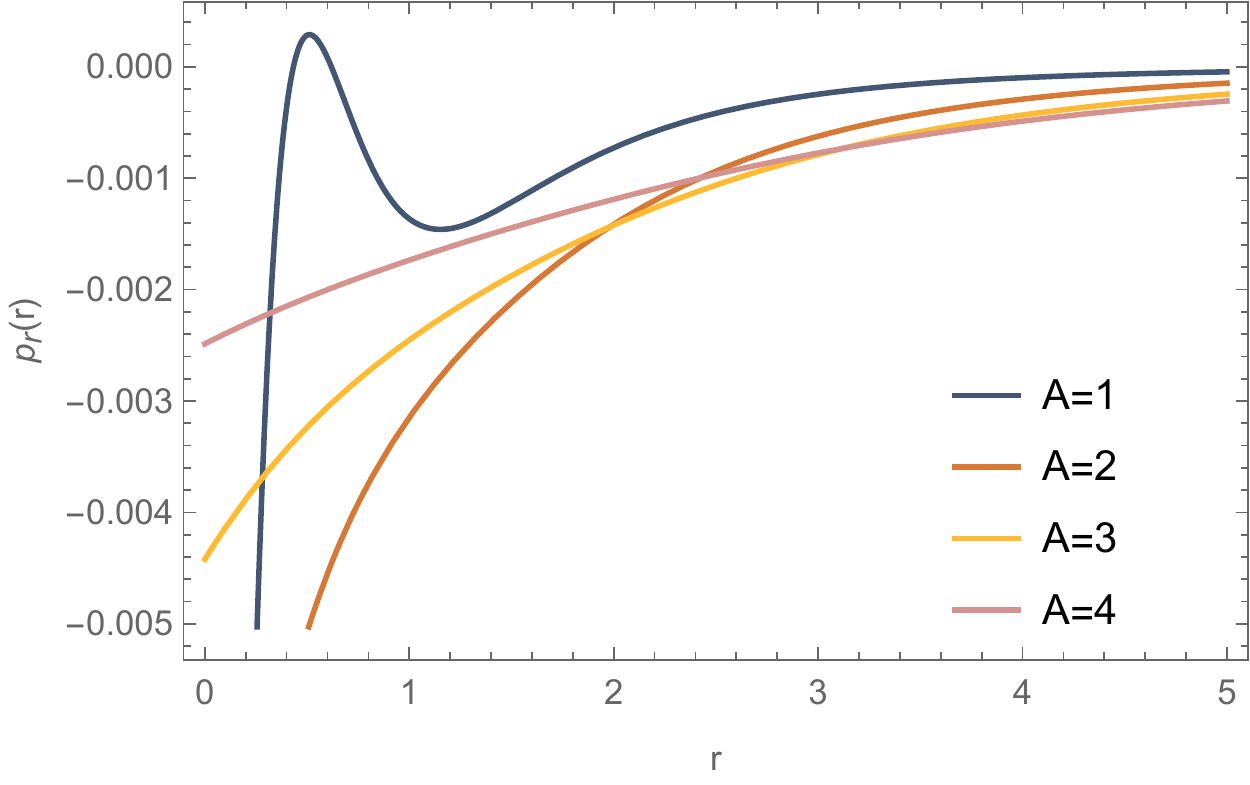}
\includegraphics[width=.40\textwidth]{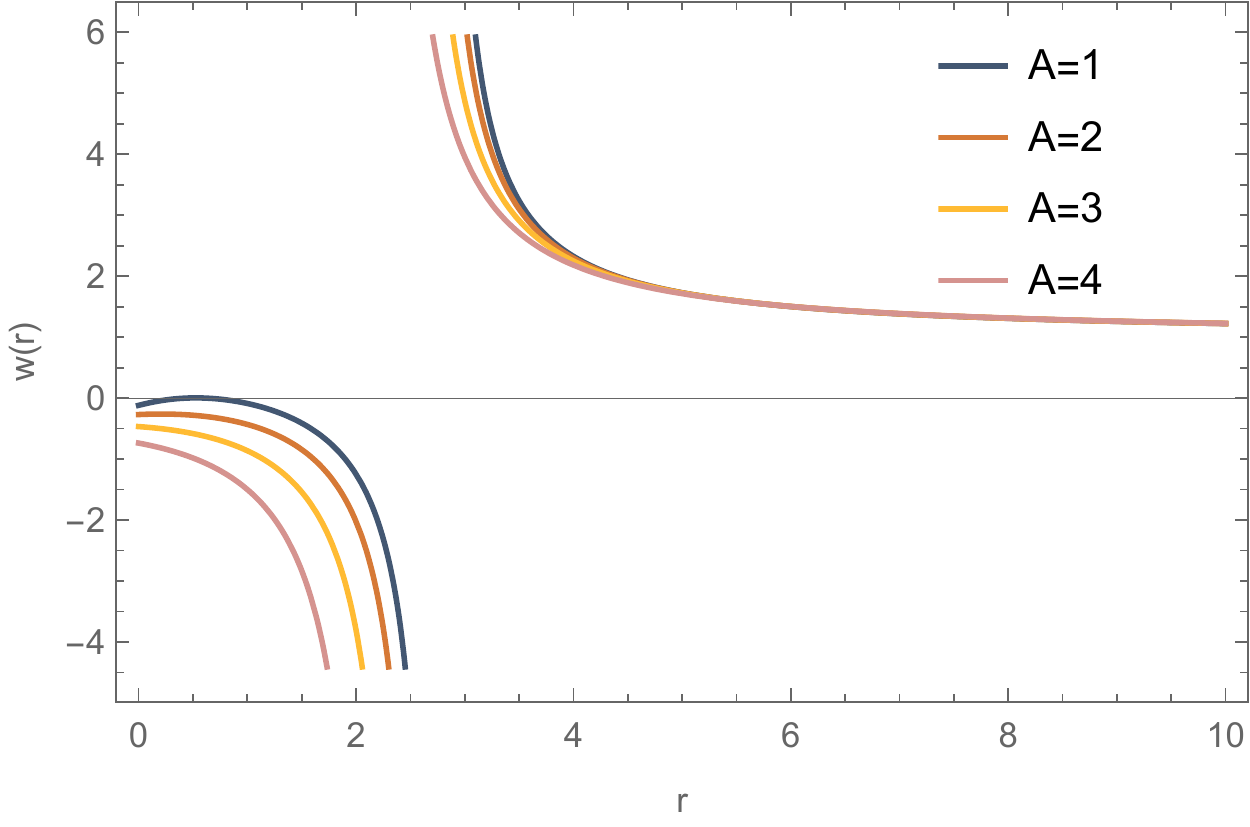}
\includegraphics[width=.40\textwidth]{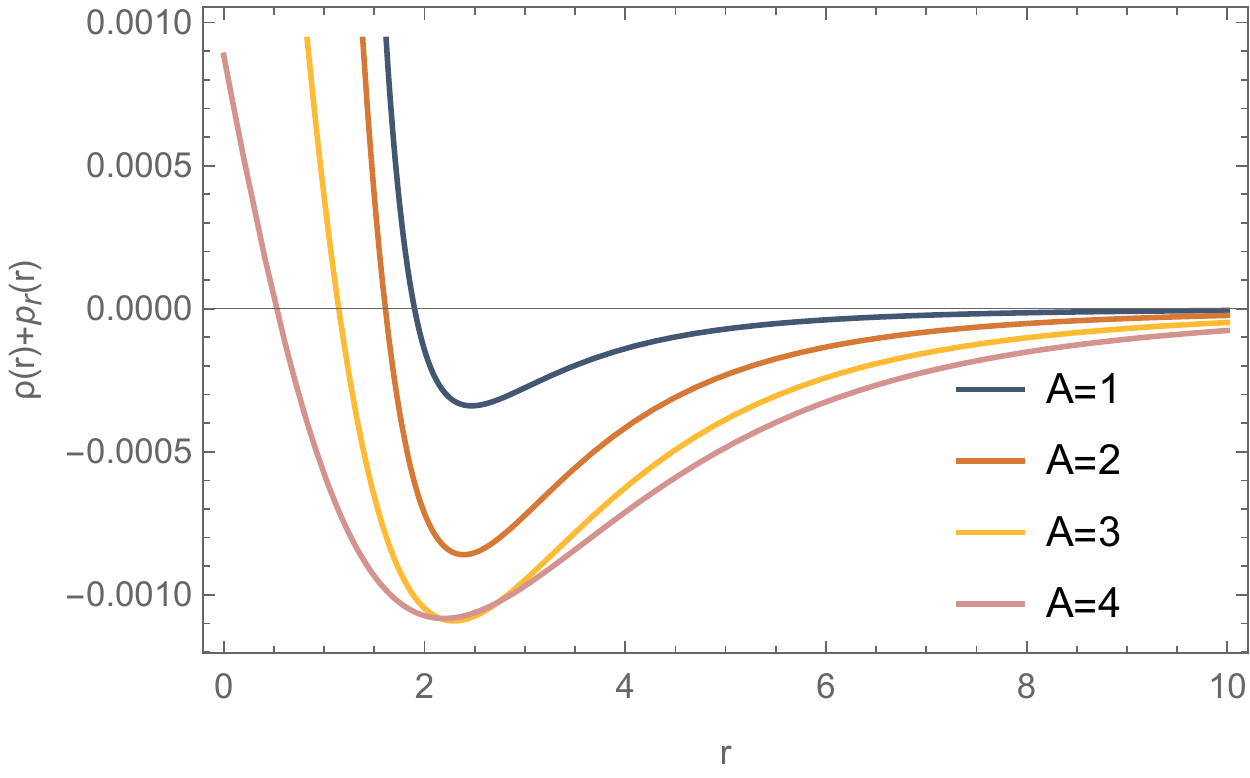}
\caption{The energy density $\rho(r)$, the radial pressure $p_r(r)$, the equation of state $w(r)$ and the sum of the energy density and pressure $\rho(r)+p_r(r)$ for $m=1$, while varying the scalar length scale $A$ for the four dimensional case.}
\label{matterD444}
\end{figure}

Now setting $m=-c_2/6$ the potential reads
\begin{equation} \label{d4pot}
V(\phi) = -\frac{3 m \Big(8 \sqrt{\pi } \phi +3 \sin \left(4 \sqrt{\pi } \phi \right)+\left(\pi -4 \sqrt{\pi } \phi \right) \cos \left(4 \sqrt{\pi } \phi \right)-2 \pi \Big)}{16 \pi  A^3}~,
\end{equation}
while its asymptotic behavior reads
\begin{equation} V(r\to\infty) \sim \frac{A^2 m}{10 \pi  r^5}-\frac{13 \left(A^4 m\right)}{70 \pi  r^7}+\frac{9 A^6 m}{35 \pi  r^9}+\mathcal{O}\left(\left(\frac{1}{r}\right)^{11}\right)~.\end{equation}
Of course for negligible scalar charge, we obtain the Schwarzschild black hole
\begin{equation}b(r,A\to0) \sim \left(1-\frac{2 m}{r}\right)+\frac{2 A^2 m}{5 r^3}+\mathcal{O}\left(A^4\right)~,
\end{equation}
and for vanishing mass $m$, we obtain pure Minkowski spacetime, i.e $b(r)=1$, hence $m$ is a genuine and independent scale of the solution, hence the theory cannot be mapped conformally to a non-minimally coupled theory in the Einstein frame.
The horizon is obtained by solving $b(r)=0$, however we cannot solve analytically this equation. Near the origin $b(r)$ behaves as
\begin{equation} b(r\to0)\sim \frac{4 A-6 \pi  m}{4 A}+\frac{6 m r}{A^2}-\frac{3 (\pi  m) r^2}{2 A^3}+\mathcal{O}\left(r^3\right)~,\end{equation}
where the term that dominates is the constant term. In order for $b(r)$ to have a root in the region $0\le r< \infty$, we need to make the constant term negative, since the $\mathcal{O}(r)$ term is positive. Hence, the scalar charge $A$ provides a bound for the mass $m$ of the black hole and the existence of a horizon
\begin{equation}\label{mAbound}
m>\frac{2 A}{3 \pi }~.\end{equation}
\begin{figure}[h]
\centering
\includegraphics[width=.40\textwidth]{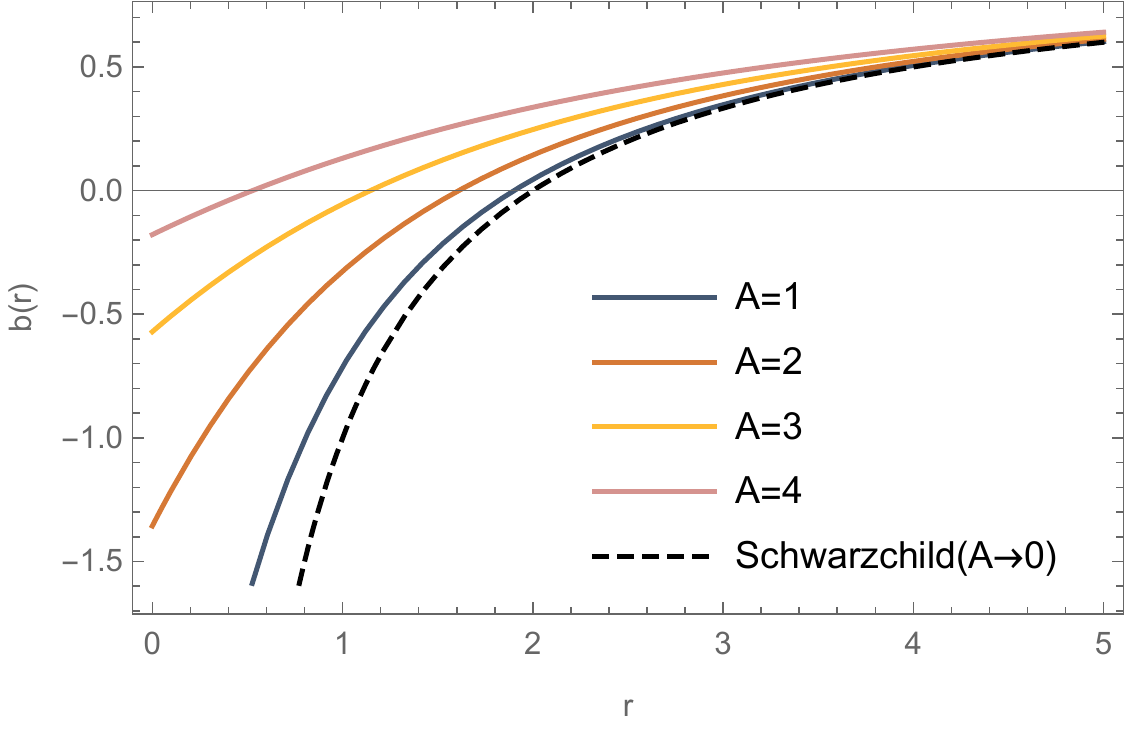}
\includegraphics[width=.40\textwidth]{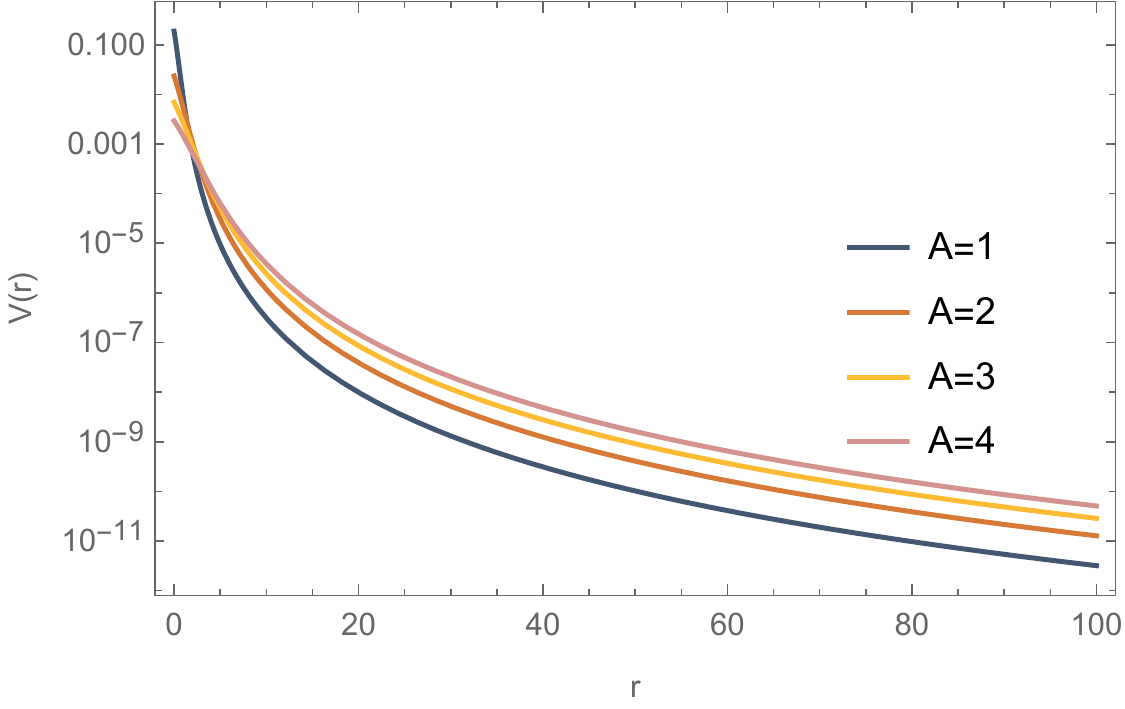}
\includegraphics[width=.40\textwidth]{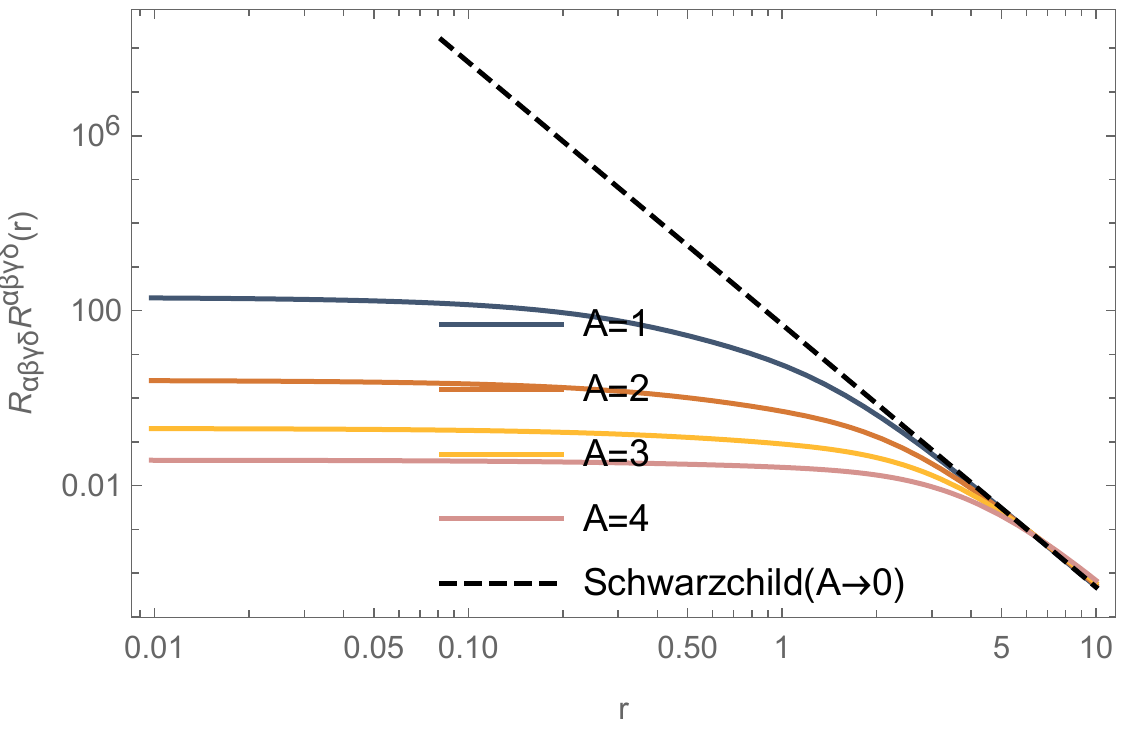}
\includegraphics[width=.40\textwidth]{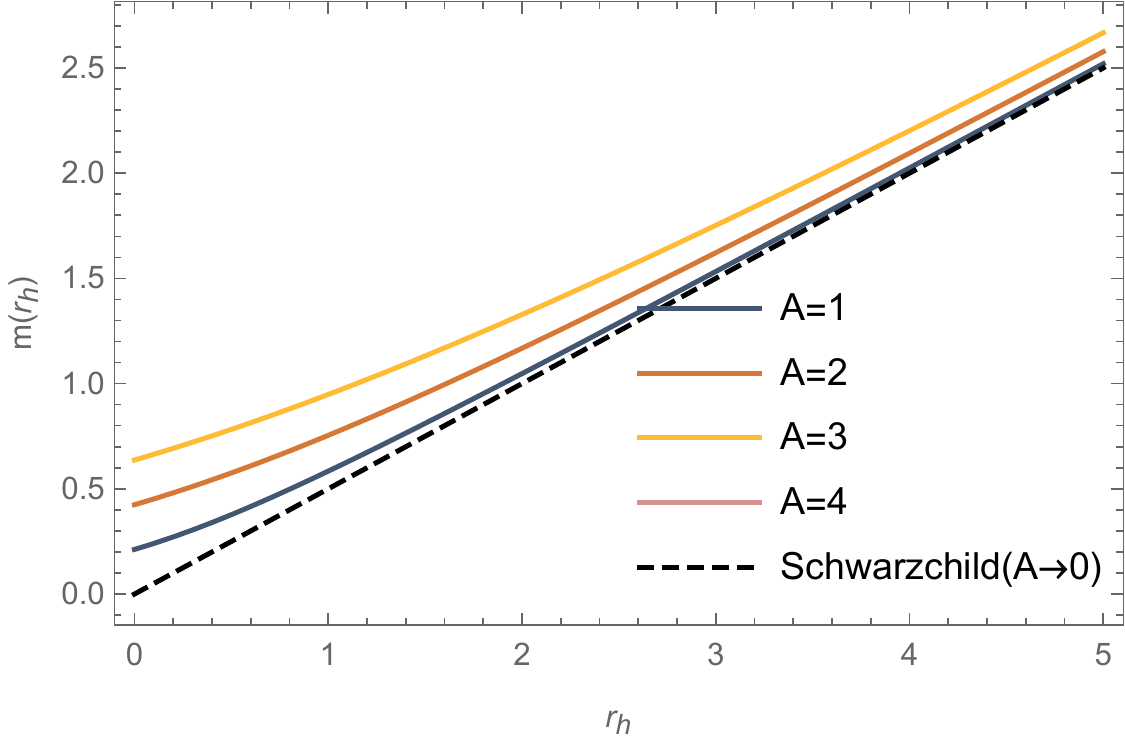}
\includegraphics[width=.40\textwidth]{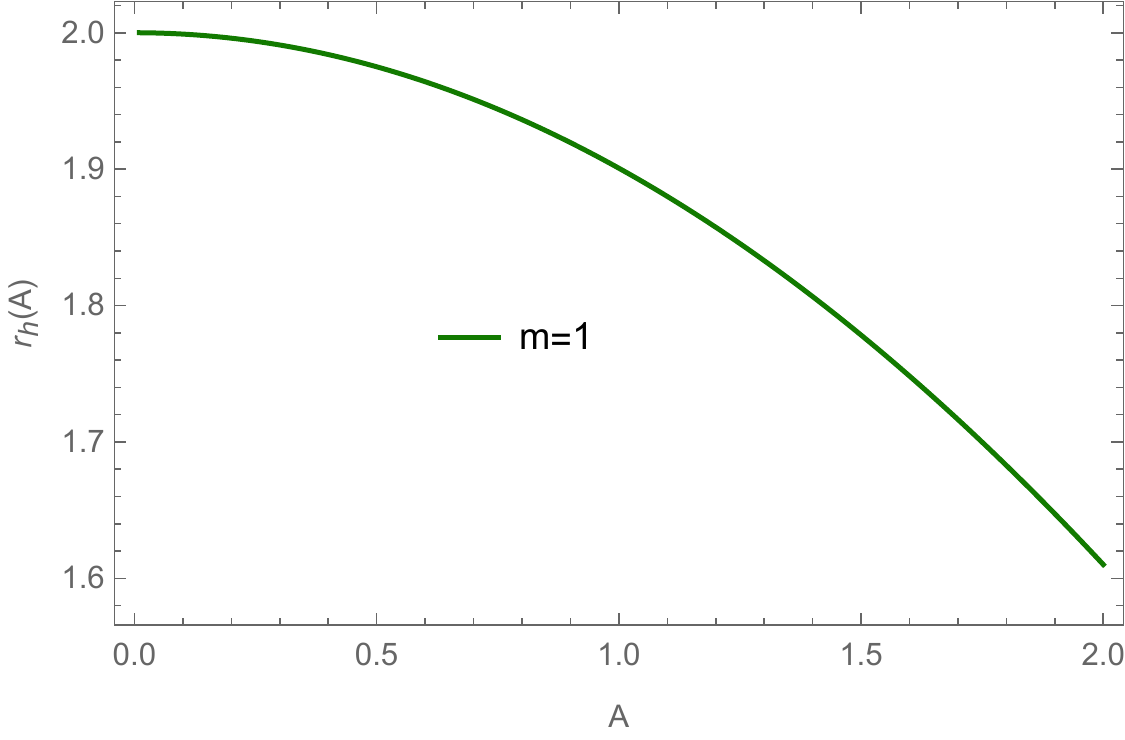}
\includegraphics[width=.40\textwidth]{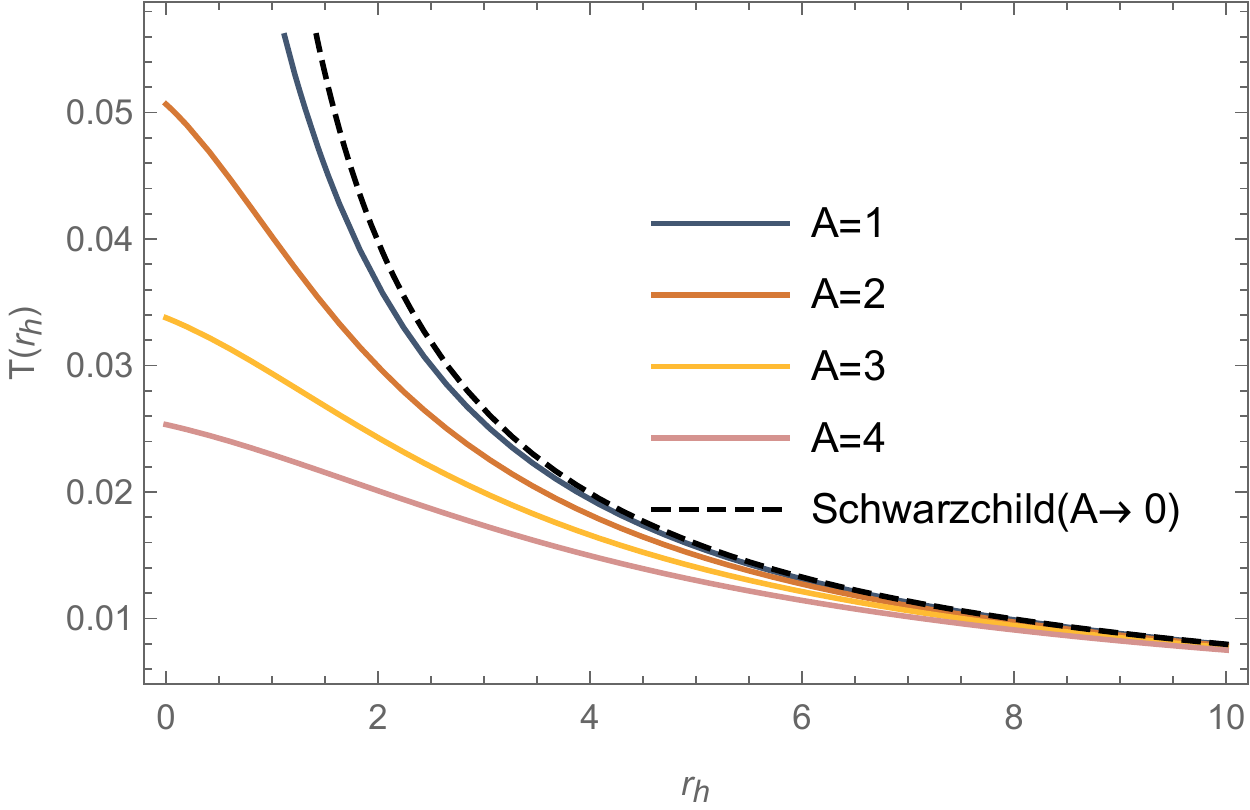}
\caption{Plots of the metric function $b(r)$, the potential $V(r)$ and the square of the Riemann tensor as functions of the radial distance $r$, the mass as function of the event horizon radius, the horizon radius as a function of the scalar charge $A$ and the temperature $T(r_h)$ of the black hole as function of the horizon radius.}
\label{D444}
\end{figure}
In Fig.~\Ref{D444} we plot the metric function $b(r)$, the potential, the Kretschmann scalar and the mass of the black hole for different values of the scalar charge $A$, the radius of the event horizon as a function of the scalar charge for a fixed value of mass, that satisfies the mass bound \eqref{mAbound}  (note that the scalar charge is equal to or larger than the black hole mass in these plots, however, the effect of the scalar hair on the black hole spacetime is negligible and small values of $A$ in comparison to the mass basically correspond to the Schwarzchild black hole), and the temperature of the black hole as a function of the horizon radius (we calculated the temperature as in the $D=3$ case and then substituted $R_h\to\sqrt{r_h^2+A^2}$ to plot as a function of $r_h$), for several values of the scalar charge $A$. We can see that, as the phantom matter is getting stronger, the black hole shrinks in size and hence the phantom black hole will not be thermodynamically preffered, since it has a smaller horizon radius. For $A\to 0$, we obtain, as expected, the horizon radius of the Schwarzschild black hole. Regarding the thermal stability of the black hole in the canonical ensemble we can see that the temperature is getting smaller as the black hole is getting larger, i.e $T'(r_h)<0$ and the mass of the black hole always increases hence $m'(r_h)>0$ so the heat capacity of the black hole will be negative indicating the thermodynamically locally unstable nature of the spacetime, which may occur in asymptotically flat spacetimes due to $T'(r_h)<0$ as $r_h$ grows. In Fig. \ref{heat4} we depict this exact behaviour alongside the heat capacity of the Schwarzschild black hole which also suffers from this instability \cite{Altamirano:2014tva}.
The entropy of the black hole may be computed using Wald formula as in the previous section only to yield the Bekenstein-Hawking area law
\begin{equation} \mathcal{S}(r_h) = \frac{\mathcal{A}}{4G}~,\end{equation}
where $\mathcal{A} = 4\pi W(r_h)^2$ is the area of the black hole.
Regarding the nature of the matter threading the black hole spacetime, we calculate the energy density, the radial pressure, their sum and the equation of state
\begin{eqnarray}
&&\rho(r) = \frac{-9 \pi  A^2 r^2+18 r^2 \left(A^2+r^2\right) \tan ^{-1}\left(\frac{r}{A}\right)+12 A^3 r-2 A^5+18 A r^3-9 \pi  r^4}{16 \pi  A^3 \left(A^2+r^2\right)^2}~, \nonumber \\
&& p_r(r) = \frac{-9 \pi  A^2 r^2+18 r^2 \left(A^2+r^2\right) \tan ^{-1}\left(\frac{r}{A}\right)+12 A^3 r-2 A^5+18 A r^3-9 \pi  r^4}{16 \pi  A^3 \left(A^2+r^2\right)^2}~, \nonumber \\
&& \rho(r) + p_r(r) = \frac{-6 \left(A^2+r^2\right) \tan ^{-1}\left(\frac{r}{A}\right)-2 A^3+3 \pi  A^2-6 A r+3 \pi  r^2}{8 \pi  A \left(A^2+r^2\right)^2}~, \nonumber \\
&& w(r) = \frac{9 \pi  A^2 r^2-18 r^2 \left(A^2+r^2\right) \tan ^{-1}\left(\frac{r}{A}\right)-12 A^3 r+2 A^5-18 A r^3+9 \pi  r^4}{-15 \pi  A^2 r^2+6 \left(A^2+r^2\right) \left(2 A^2+3 r^2\right) \tan ^{-1}\left(\frac{r}{A}\right)+24 A^3 r+2 A^5-6 \pi  A^4+18 A r^3-9 \pi  r^4}~,
\end{eqnarray}
which are plotted in Fig. \ref{matterD444}. As stated in Section \ref{sec:buildBH} the NEC is satisfied inside and at the event horizon of the black hole and violated for any $r>r_h$ due to the phantom nature of the scalar field. Also, $w(r_h)=-1$ as we already discussed.
As we can see from Fig. \ref{matterD444} the WEC is violated for a region $r>r_h$ and holds for $r\leq r_h$. The equation of state differs from the $2+1$-dimensional case and we attribute this behaviour to the asymptotically flat nature of spacetime that is encoded in $\rho,p_r$ through $b(r),V(r)$.

As we discussed in the $D=3$ case, we impose the relation
\begin{align}\label{cond4}
\frac{m}{A^3} = |\chi|\, G^{-2}\,,
\end{align}
with $\chi \in \mathbb R$ a real dimensionless parameter, between the mass of the black hole and the scalar charge which makes
the potential \eqref{d4pot} independent of these parameters, which now will read
\begin{equation} \label{d4pot1}
V(\phi) = -\frac{3}{16 \pi G^2}|\chi| \Big(8 \sqrt{\pi } \phi +3 \sin \left(4 \sqrt{\pi } \phi \right)+\left(\pi -4 \sqrt{\pi } \phi \right) \cos \left(4 \sqrt{\pi } \phi \right)-2 \pi \Big) ~.
\end{equation}

Let us now examine the conditions for thermal equilibrium in such a case.
To this end, we assume the validity of the standard first law of Thermodynamics. In this case, one would expect that the extremality condition \eqref{cond4} would
 correspond to a zero temperature. However, let us see how close to thermal equilibrium the $D=4$ compact object is.

We first remark that the mass function at the horizon is given by
$$M(r_h) = \frac{2 A^3}{3 \left(-2 \left(A^2+r_h^2\right) \tan ^{-1}\left(\frac{r_h}{A}\right)+\pi  A^2-2 A r_h+\pi  r_h^2\right)}~,$$
obtained from $b(r_h)=0 \to M = M(r_h)$.
The temperature
$$T(r_h) =\frac{2 r_h \tan ^{-1}\left(\frac{r_h}{A}\right)+2 A-\pi  r_h}{2 \pi  \left(-2 \left(A^2+r_h^2\right) \tan ^{-1}\left(\frac{r_h}{A}\right)+\pi  A^2-2 A r_h+\pi  r_h^2\right)}~,$$
while the entropy is
$$S(r_h) = \pi  \left(A^2+r_h^2\right)~.$$
The first law of thermodynamics is expressed as $dM = TdS \to M'(r_h)dr_h = T(r_h)S'(r_h)dr_h$, provided the quantities that enter this formula are non zero.
In our $D=4$ case, we have
\begin{align}\label{dM}
dM = M'(r_h)dr_h =\frac{4 A^3 \left(2 A-r_h \left(\pi -2 \tan ^{-1}\left(\frac{r_h}{A}\right)\right)\right)}{3 \left(A^2 \left(\pi -2 \tan ^{-1}\left(\frac{r_h}{A}\right)\right)-2 A r_h+r_h^2 \left(\pi -2 \tan ^{-1}\left(\frac{r_h}{A}\right)\right)\right){}^2} dr_h~,
\end{align}
and
\begin{align}\label{Tdef}
T(r_h)S'(r_h)dr_h =\frac{r_h \left(2 r_h \tan ^{-1}\left(\frac{r_h}{A}\right)+2 A-\pi  r_h\right)}{-2 \left(A^2+r_h^2\right) \tan ^{-1}\left(\frac{r_h}{A}\right)+\pi  A^2-2 A r_h+\pi  r_h^2}dr_h~,
\end{align}
which shows that in general $dM \neq TdS$  unless $dM=T=0$. The latter condition would be guaranteed in a self consistent way from \eqref{dM} and \eqref{Tdef}, provided
\begin{align}\label{condhor}
2 A-r_h \left(\pi -2 \tan ^{-1}\left(\frac{r_h}{A}\right)\right) =0
\quad  \Rightarrow \quad \frac{A}{r_h} =  \left(\frac{\pi}{2}  -  \tan ^{-1}\left(\frac{r_h}{A}\right)\right)\,.
\end{align}
This is consistent with the discrete values the mass and charges are allowed to take on in this case.

The relation \eqref{condhor} implies as a fixed relation of the ratio of the horizon radius to the scalar charge
\begin{align}\label{rhA}
\frac{r_h}{A} = \zeta_0\, \to  \, +\infty\,.
\end{align}

Ignoring for the moment the singular limit \eqref{rhA}, we may combine \eqref{rhA} with the definition of the horizon  radius $b(r=r_h)=0$ ({\it cf.} \eqref{bhor}), we obtain (using \eqref{cond4}, \eqref{condhor} and \eqref{d4mass})
\begin{align}\label{finalcvalm}
|\chi | \, G^{-2} = \frac{\zeta_0}{3\, A^2} = \frac{r_h}{3\, A^3}\,,
\end{align}
From \eqref{cond4} this implies
\begin{align}\label{mA}
G\, m = \frac{r_h}{3} = \frac{\zeta_0}{3} \, A~,
\end{align}
which is the condition for a self consistent solution in $D=4$ with secondary hair.

We note that equation \eqref{condhor}, although formally valid in the singular limit $\zeta_0 \to +\infty$, nonetheless it is valid to a {\it good approximation} for $r_h/A \gtrsim \mathcal O(100)$ (for such values eq.~\eqref{condhor} is valid to an accuracy $10^{-7}$ and the accuracy is increasing  rapidly for higher values, as we approach $r_h/A \to \infty$). Then  we may interpret such results as implying that for relatively heavy compact objects {\it i.e.} with very large masses as compared to the magnitude of the scalar charge $S=A$, one may have {\it approximately only} thermal equilibrium, which is heavily violated as the ratio $m/A $ decreases.

As we have already discussed in the $D=3$ case, in the first law of thermodynamics one has to consider that $E\neq m$, due to the fact that the Lagrangian of the matter fields is not entirely independent of the black hole mass \cite{Ma:2014qma}. The above-described analysis, leading to the singular limit \eqref{mA}, \eqref{rhA}, confirms that the situation characterising the $D=3$ case is also valid qualitatively in the $D=4$ case, that is, here we also have $E \ne m$  unless the secondary-hair charge $A \to 0$. Nonetheless the solution in the $D=4$ case is complicated, which prevents us from performing simple calculations in order to make it explicit that the internal energy is not equal to the conserved black hole mass due to the presence of the secondary-hair charge $A$. 

\subsection{Higher Dimensions}

In what follows we shall describe regular black hole solutions in $D>4$ spacetime dimensions. For concreteness and brevity we shall only deal explicitly with three cases $D=5,6$ and $10$. We recall
that gravitational-Schwarzchild mass term will be given by the
\begin{align}\label{bmass}
{\rm Schwarzschild-mass} \propto  \,\, \mathcal{O}\left(\left(1/r\right)^{D-3}\right) \, \in  \, b(r \to \infty) \,,
\end{align}
term of the limiting expression for the metric function $b(r)$ at infinity, $ r \to \infty$. As we shall see below, the spacetime for higher dimensions can be asymptotically flat, but not $asymptotically~ free$, since the scalar field has a fixed fall-off for any spacetime dimension. This is going to affect the definition of mass and the corresponding scalar potential.

\subsubsection{The $D=5$ case}
For $D=5$, the asymptotically flat solution corresponds to
\begin{multline}
b(r)=\frac{1}{3 A^4}
\Bigg(c_2 \left(A^2 \left(2 \sqrt{A^2+r^2}-3 r\right)+2 r^2 \left(\sqrt{A^2+r^2}-r\right)\right)+\\
2 A^2 \left(A^2 \sqrt{A^2+r^2}+\left(A^2+r^2\right)^{3/2} \ln \left(4 \left(A^2+r^2\right)\right)-r \left(3 A^2+2 r^2\right) \ln \left(\sqrt{A^2+r^2}+r\right)\right) \sqrt{A^2+r^2}\Bigg)~,
\end{multline}
where we have adjusted $c_1$ in order to make the vacuum energy vanish.
The asymptotic expressions of the metric function read
\begin{eqnarray}\label{blarger}
&&b(r\to0)\sim\frac{2}{3} \left(\frac{c_2}{A^2}+2 \ln (A)+1+\ln (4)\right)-\frac{r \left(2 A^2 \ln (A)+c_2\right)}{A^3}+\mathcal{O}\left(r^2\right)~,\\
&&b(r\to\infty) \sim 1+\frac{4 A^2 \ln (r)+A^2+4 A^2 \ln (2)+2 c_2}{8 r^2}+\mathcal{O}\left(\left(\frac{1}{r}\right)^4\right)~.
\end{eqnarray}
One can check that a computation of the Komar mass (\ref{komar}) will yield a divergent conserved black hole mass, if one takes the hypersphere on which the integral is evaluated to have an infinite radius. To bypass this problem we will define a cutoff radius $R_{\text{cutoff}}$ with $R_{\text{cutoff}} \gg A$. Then, one can read off the mass of the black hole as
\begin{equation} m = \frac{1}{16} \left(-4 A^2 \ln \left(2 R_{\text{cutoff}}\right)+A^2-2 c_2\right)~,\end{equation}
which it is clear that will reduce the the Schwarzchild mass when $A\to0$. This in turn implies that the theory holds up to a critical energy scale $E_{\text{cutoff}}$. The scalar potential turns out to contain the black hole mass, as well as the cutoff radius (due to its complexity we will not give the explicit expression). As a result the produced compact object might have a problematic thermodynamic prescription, since the mass is not allowed to vary, but is instead fixed by the theory.

For completion we mention that, for small $A/r$,  we obtain the Tangherlini-Schwarzschild black hole at zeroth order as expected \cite{Tangherlini:1963bw}
\begin{equation} b(r,A/r \ll 1) \sim \left(1+\frac{c_2}{4 r^2}\right)+\frac{A^2 \left(3 r^2 (4 \ln (r)+1+\ln (16))-4 c_2\right)}{24 r^4}+O\left(A^3\right)~.
\end{equation}

\begin{figure}[h]
\centering
\includegraphics[width=.40\textwidth]{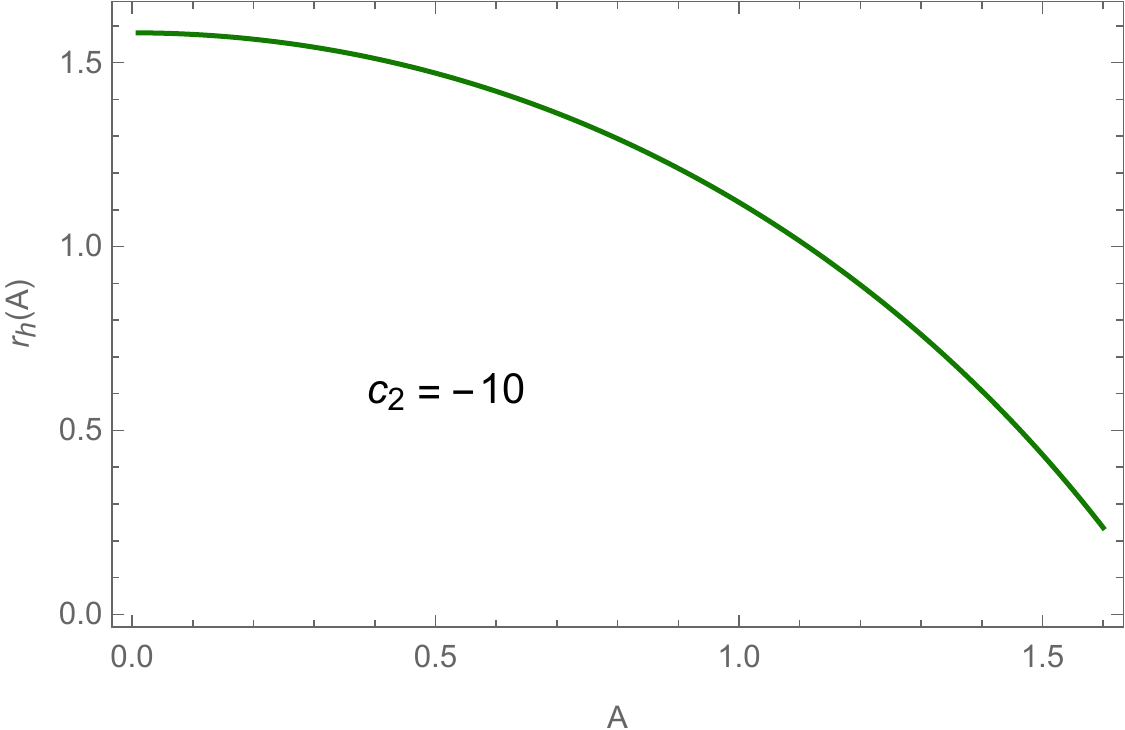}
\includegraphics[width=.40\textwidth]{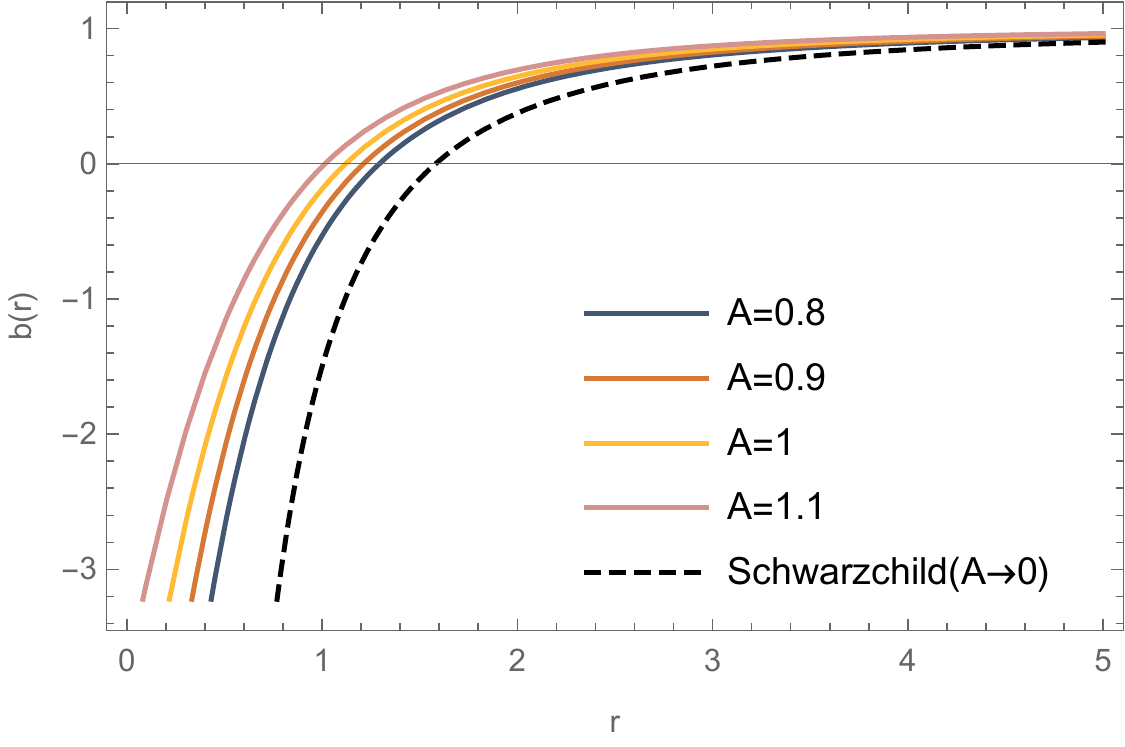}
\includegraphics[width=.40\textwidth]{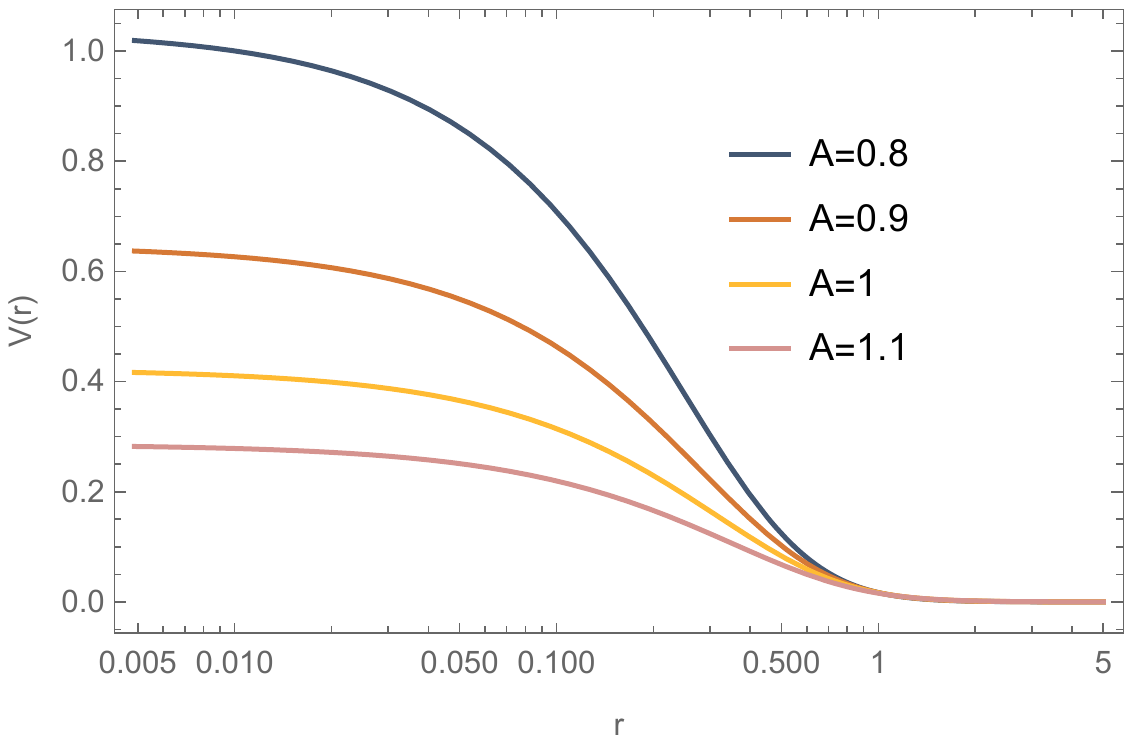}
\includegraphics[width=.40\textwidth]{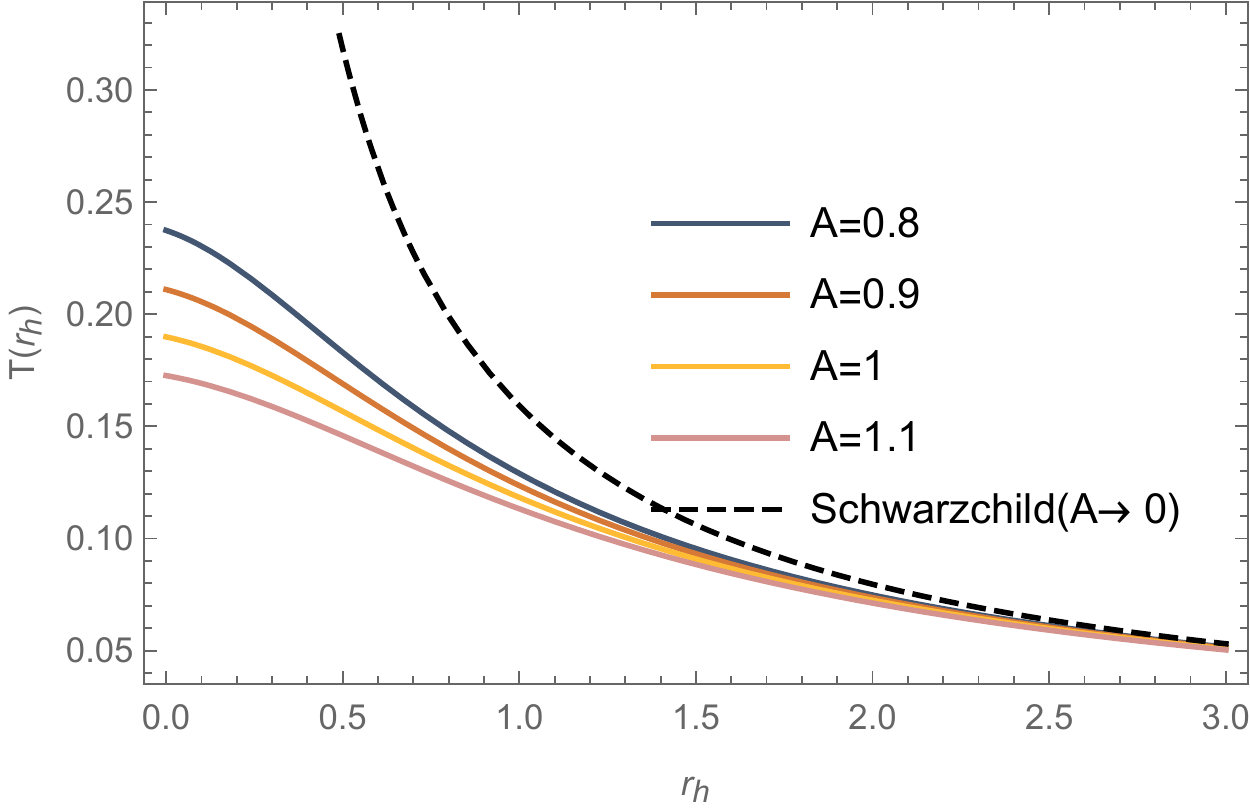}
\caption{The horizon radius as a function of $A$, the metric function $b(r)$ the potential $V(r)$ and the temperature $T(r_h)$ for $c_2=-10$ while changing $A$, for the five-dimensional black hole.}
\label{d5}
\end{figure}
\begin{figure}[h]
\centering
\includegraphics[width=.40\textwidth]{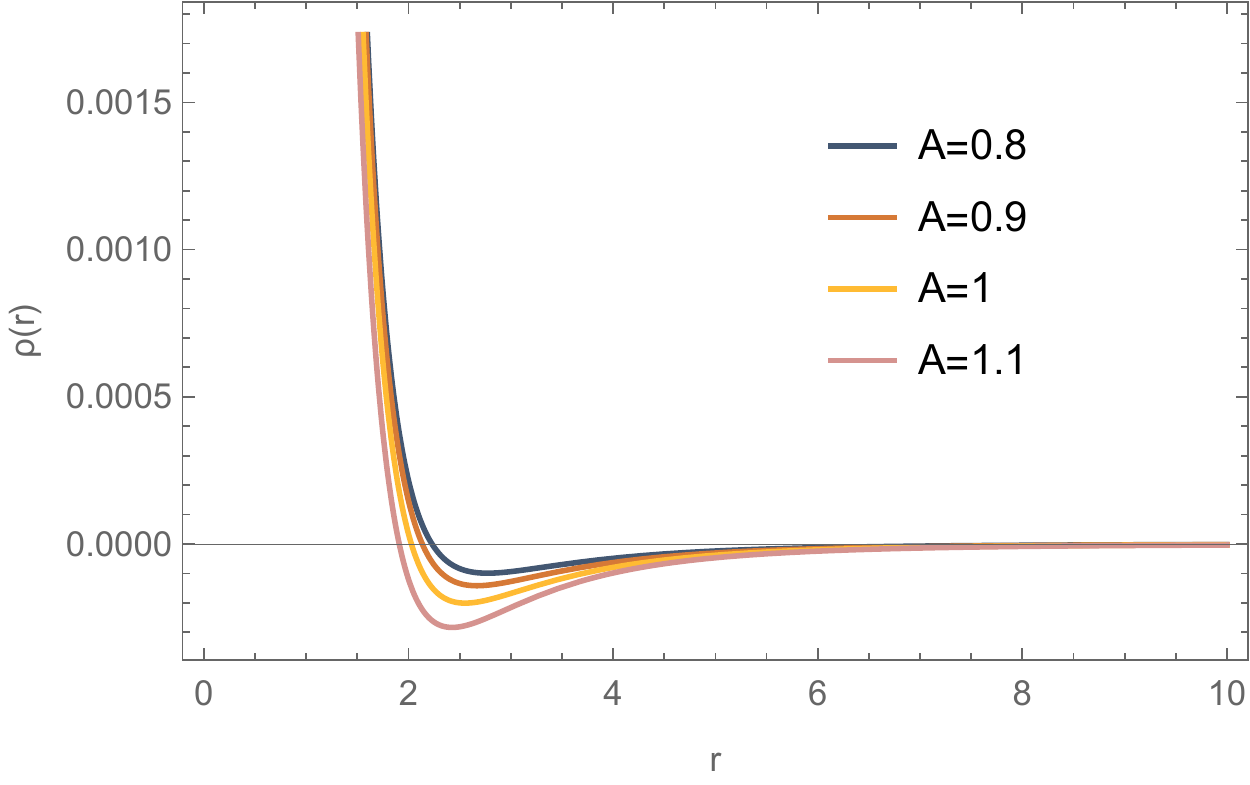}
\includegraphics[width=.40\textwidth]{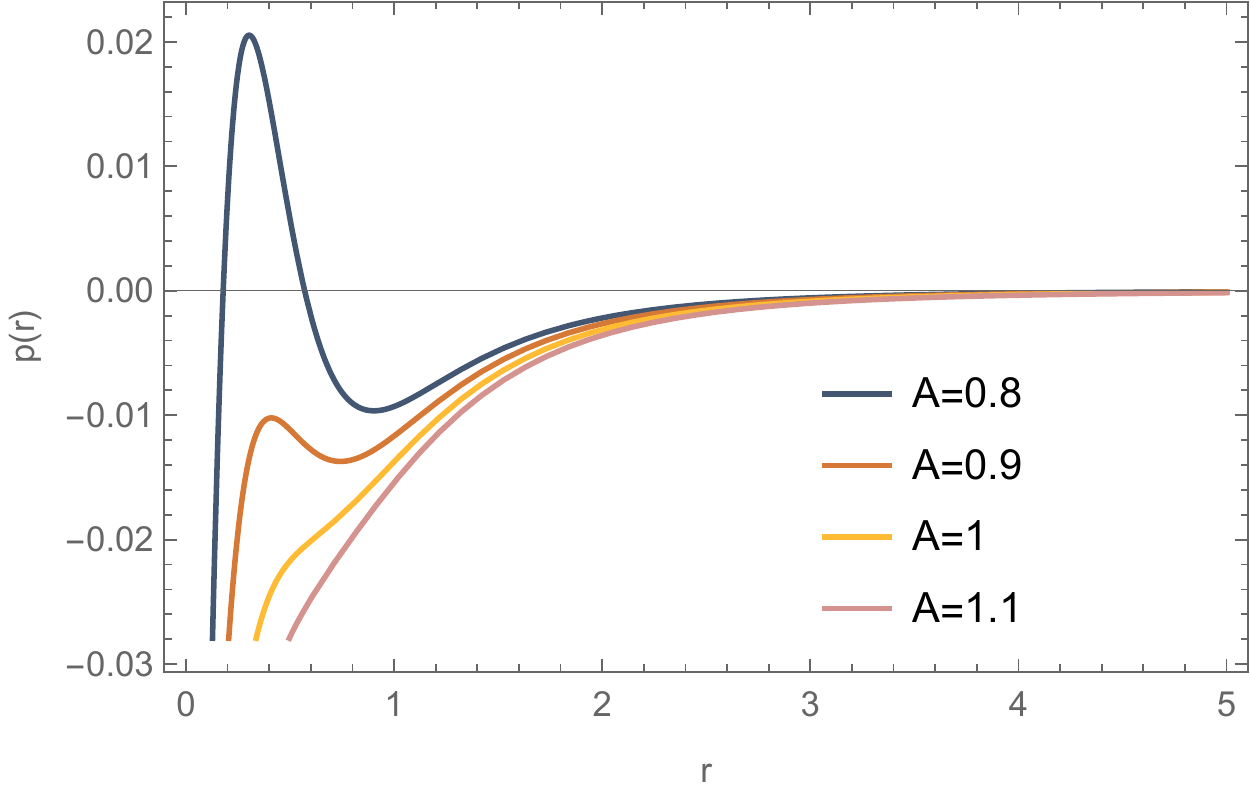}
\includegraphics[width=.40\textwidth]{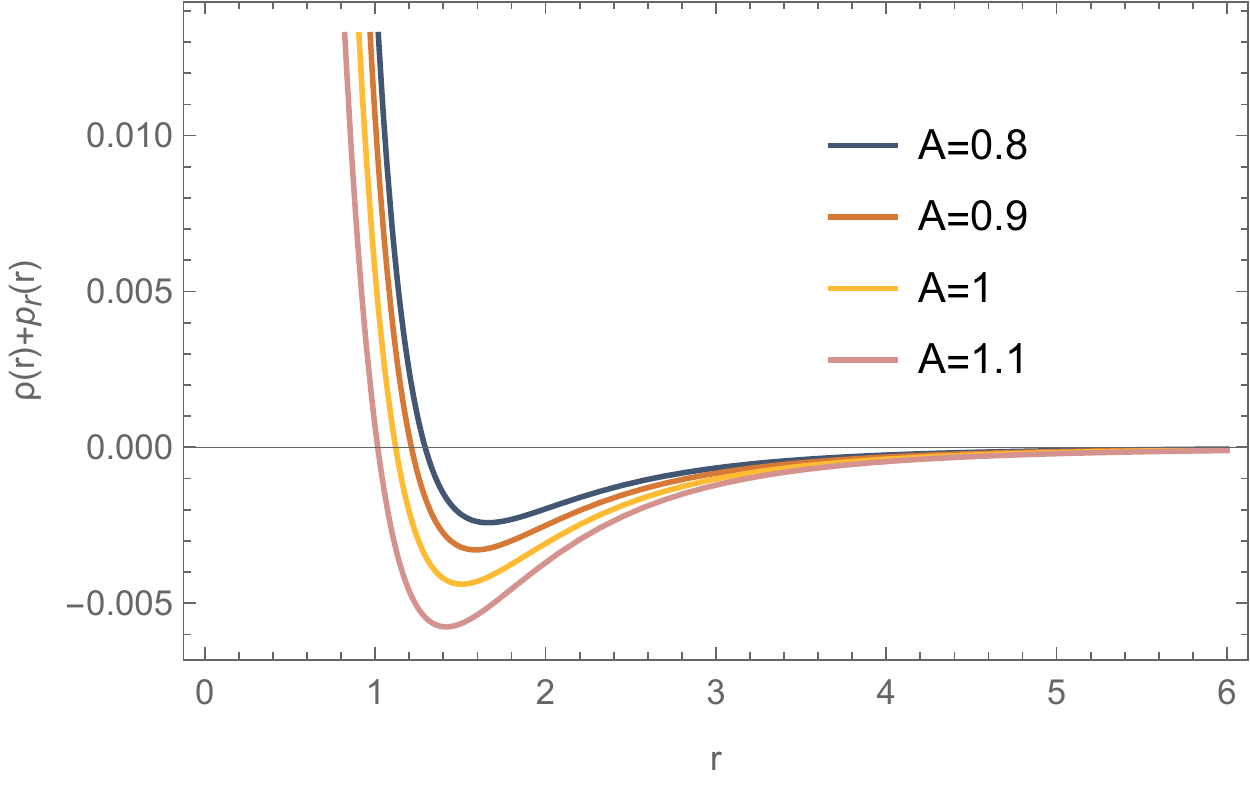}
\includegraphics[width=.40\textwidth]{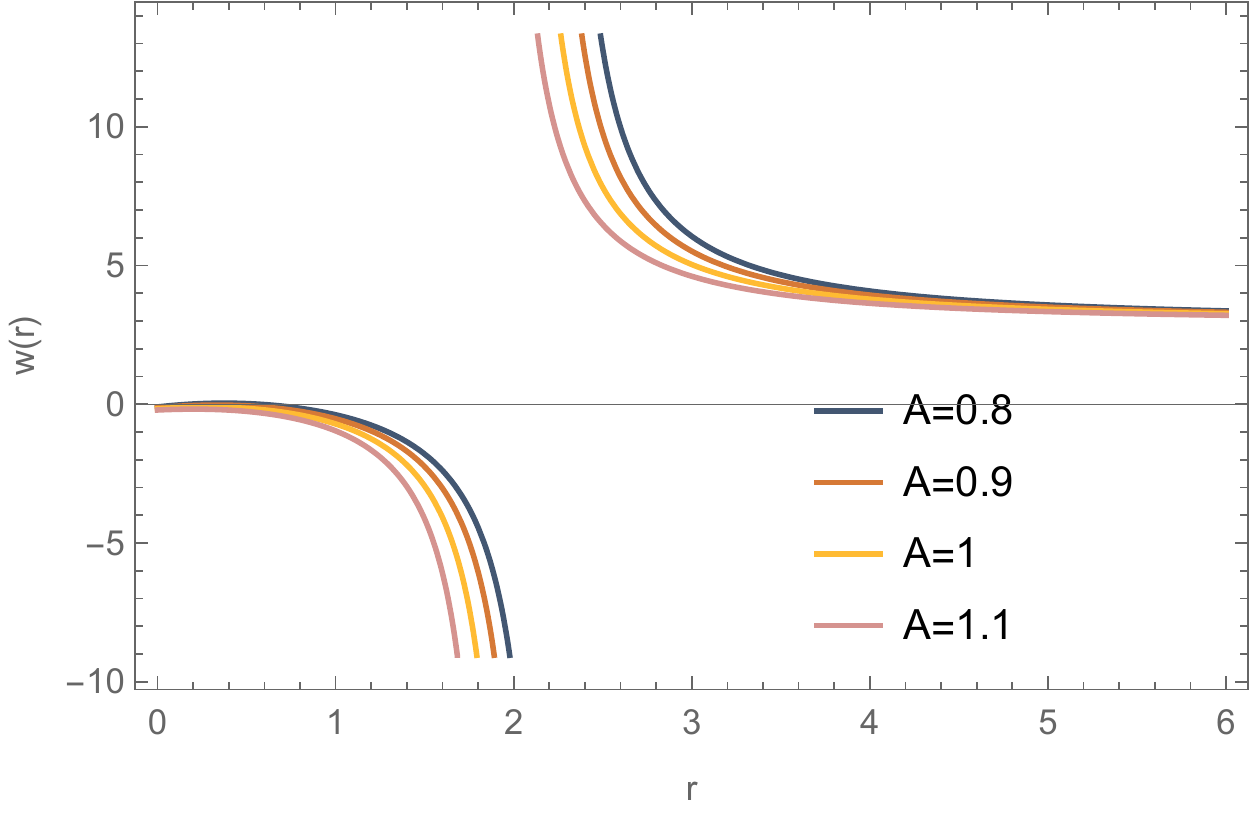}
\caption{The energy density, the radial pressure, their sum and the equation of state are plotted while changing $A$ for the five dimensional case.}
\label{matter5}
\end{figure}

Since we have performed detailed analyses for the cases $D=3,4$, in what follows, in order to discuss the resulting physics, we will not show any calculations (which are similar to the previously studied lower-dimensional cases), but only plot the relevant quantities. Specifically, In Fig. \ref{d5} we plot the horizon radius, the metric function $b(r)$, the potential and the temperature for $T(r_h)$ for the five dimensional black hole. For the computation of temperature, we at first computed it as in the $D=3$ case and then substituted $R_h\to\sqrt{r_h^2+A^2}$ in the final result to express it as a function of $r_h$. We can see that the horizon radius is getting smaller as the scalar charge is getting stronger in agreement with the four dimensional case. The potential and temperature also behave similar to the four dimensional case, being always positive. In Fig. \ref{matter5} we plot the quantities related to the matter threading the black hole. The energy density is positive near the origin and changes sign after the black hole horizon, indicating in this way a divergence in the equation of state and the violation of WEC. The radial pressure develops one global maximun and a local minima which can be numerically evaluated but nothing special happens at these points. The NEC holds inside and at the horizon of the black hole while violated for any $r>r_h$ and also $w(r_h)=-1$, which are general features of the theory at hand as we have already discussed. Comparing with the four dimensional case, we can see that besides the fact that the radial pressure might be positive, the matter sector, behaves in a similar manner regarding the energy conditions and their violation.

%\section{The asymptotically flat solution for $D>5$}

\subsubsection{The $D=6$ case}

For $D=6$ we find
\begin{eqnarray}
&&b(r)=\frac{6 \pi  A^2 c_2 r^2-6 c_2 \left(A^2+r^2\right)^2 \tan ^{-1}\left(r/A\right)-10 A^3 c_2 r+3 \pi  A^4 c_2+16 A^5 r^2+32 A^7-6 A c_2 r^3+3 \pi  c_2 r^4}{16 A^5 \left(A^2+r^2\right)}~,\\
&&b(r\to0)\sim \frac{32 A^3+3 \pi  c_2}{16 A^3}-\frac{c_2 r}{A^4}+\frac{r^2 \left(3 \pi  c_2-16 A^3\right)}{16 A^5}+\mathcal{O}\left(r^3\right)~,\\
&&b(r\to\infty) \sim 1+\frac{A^2}{r^2}+\frac{c_2}{5 r^3}-\frac{A^4}{r^4}+\mathcal{O}\left(\left(\frac{1}{r}\right)^5\right)~,
\end{eqnarray}
where, again, $c_2$ will be related to the Schwarzchild mass of the black hole.  Following the same procedure as in the $D=5$ case we can read of the mass of the black hole as
\begin{equation} m = -\frac{A^2 R_{\text{cutoff}}}{3}-\frac{c_2}{10}~.\end{equation}
Since the mass is given in terms of the scalar charge and the cutoff radius, the hair is secondary.

The scalar potential for $D=6$ yields
\begin{align}\label{d6pot}
V(\phi) &= \frac{1}{192 \pi  A^5}\left(6 A^3 \cos \left(4 \sqrt{2 \pi } \phi \right)-140 \sin \left(2 \sqrt{2 \pi } \phi \right) \left(A^2 R_{\text{cutoff}}+3 m\right)+5 \sin \left(4 \sqrt{2 \pi } \phi \right) \left(A^2 R_{\text{cutoff}}+3 m\right)+\right.
\nonumber \\
&12 \cos \left(2 \sqrt{2 \pi } \phi \right) \left(-5 A^2 \left(\pi -2 \sqrt{2 \pi } \phi \right) R_{\text{cutoff}}+2 A^3-15 m \left(\pi -2 \sqrt{2 \pi } \phi \right)\right)+18(\left(A^3+15 m \left(\pi -2 \sqrt{2 \pi } \phi \right)\right.
\nonumber \\
&\left.\left.+5 A^2 \left(\pi -2 \sqrt{2 \pi } \phi \right) R_{\text{cutoff}}\right)\right)~.
\end{align}
It is clear that the potential contains the black hole mass, the scalar charge, as well as the cutoff radius and as a result these compact objects do not have a well behaved first law of thermodynamics due to the fact that the mass is fixed
The reader is also invited to study the analysis of \cite{Feng:2013tza} for relevant discussions.

\subsubsection{The $D=10$ case}
For $D=10$ we find
\begin{multline}
b(r) = \frac{1}{3840 A^9 \left(A^2+r^2\right)^3}\Bigg(4 A^6 r^2 \left(4544 A^7+525 \pi  c_2\right)-5110 A^5 c_2 r^3+18 A^4 r^4 \left(768 A^7+175 \pi  c_2\right)-3850 A^3 c_2 r^5\\+60 A^2 r^6 \left(64 A^7+35 \pi  c_2\right)-1050 c_2 \left(A^2+r^2\right)^4 \tan ^{-1}\left(\frac{r}{A}\right)-2790 A^7 c_2 r+A^8 \left(11264 A^7+525 \pi  c_2\right)-1050 A c_2 r^7+525 \pi  c_2 r^8\Bigg)~,
\end{multline}
\begin{eqnarray}
&&b(r\to0) \sim \left(\frac{35 \pi  c_2}{256 A^7}+\frac{44}{15}\right)-\frac{c_2 r}{A^8}+r^2 \left(\frac{35 \pi  c_2}{256 A^9}-\frac{61}{15 A^2}\right)+\frac{2
   c_2 r^3}{3 A^{10}}+\mathcal{O}\left(r^4\right)~,\\
&&b(r\to\infty) \sim 1+\frac{3 A^2}{5 r^2}-\frac{A^4}{15 r^4}+\frac{A^6}{3 r^6}+\frac{c_2}{9 r^7}-\frac{7 A^8}{5 r^8}+\mathcal{O}\left(\left(\frac{1}{r}\right)^9\right)~.
\end{eqnarray}

As we have already discussed in the previous case, we can define the conserved mass up to cutoff radius, which in the case of $D=10$ is found to be
\begin{equation} m= \frac{-3 \left(20 A^4 R_{\text{cutoff}}^3+72 A^2 R_{\text{cutoff}}^5+123 A^6 R_{\text{cutoff}}\right)-140 c_2}{2520}~,\end{equation}
and the hair is secondary, since $A$ enters the definition of mass. The scalar field theory ends up containing the conserved black hole parameter, as well as the cutoff radius and the scalar charge $A$.
\begin{align}\label{d10pot}
V(\phi) &= \frac{1}{860160 \pi  A^9}\left(56(\left(16 \cos \left(2 \sqrt{\pi } \phi \right) \left(956 A^7-4725 m \left(\pi -2 \sqrt{\pi } \phi \right)\right)+16(A^7(\left(280 \cos \left(4 \sqrt{\pi } \phi \right)+20 \cos \left(6 \sqrt{\pi } \phi \right)+\right.\right.\right.
\nonumber \\
&\left.\cos \left(8 \sqrt{\pi } \phi \right)\right)+20 \left(556 A^7+4725 m \left(\pi -2 \sqrt{\pi } \phi \right)+9 m \left(-819 \sin \left(2 \sqrt{\pi } \phi \right)+70 \sin \left(4 \sqrt{\pi } \phi \right)+5 \sin \left(6 \sqrt{\pi } \phi \right)\right)\right)+
\nonumber \\
& \left.45 m \sin \left(8 \sqrt{\pi } \phi \right)\right)+3(A^2(\left(-1680 \left(\pi -2 \sqrt{\pi } \phi \right) \cos \left(2 \sqrt{\pi } \phi \right)-3276 \sin \left(2 \sqrt{\pi } \phi \right)+20(\left(105 \left(\pi -2 \sqrt{\pi } \phi \right)\right.\right.
\nonumber \\
&\left.\left.\left.+14 \sin \left(4 \sqrt{\pi } \phi \right)+\sin \left(6 \sqrt{\pi } \phi \right)\right)+\sin \left(8 \sqrt{\pi } \phi \right)\right)R_{\text{cutoff}}(20 A^2 R_{\text{cutoff}}^2+123 A^4+72 R_{\text{cutoff}}^4)\right)~.
\end{align}

\begin{figure}[h]
\centering
\includegraphics[width=.40\textwidth]{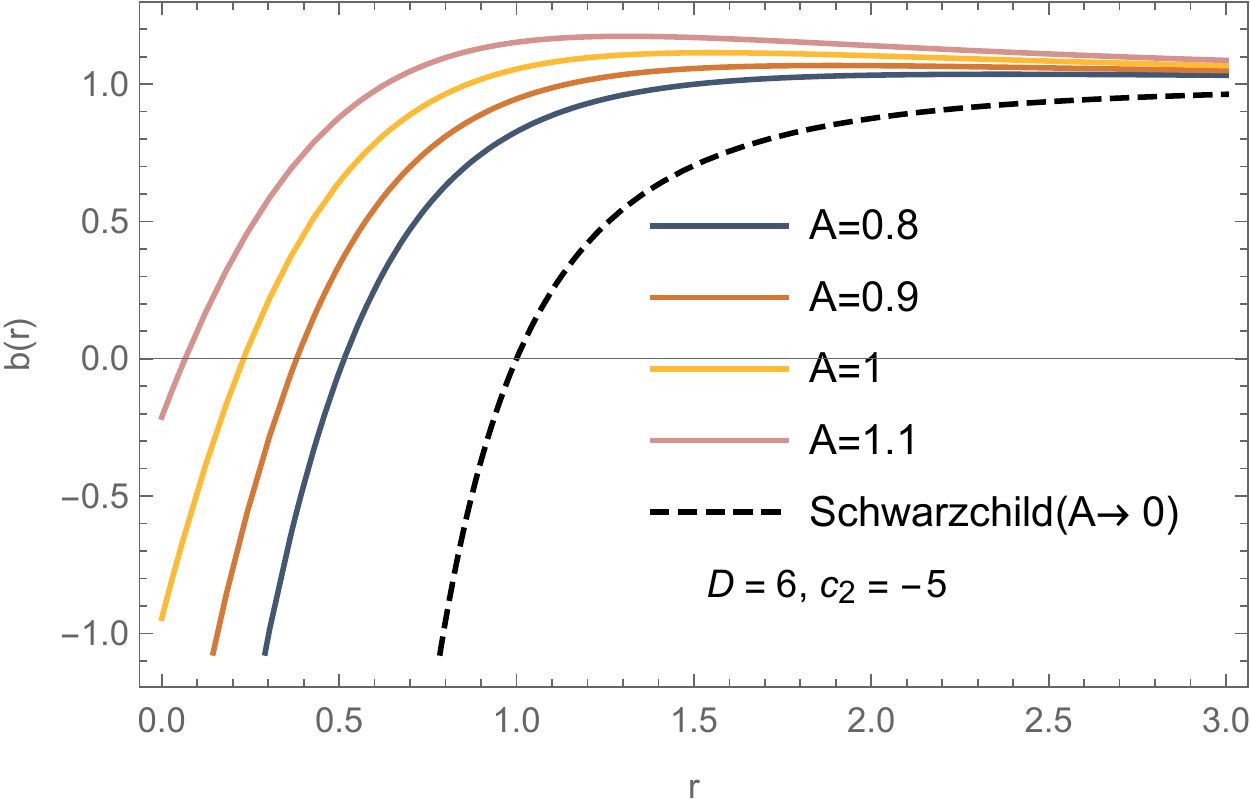}
\includegraphics[width=.40\textwidth]{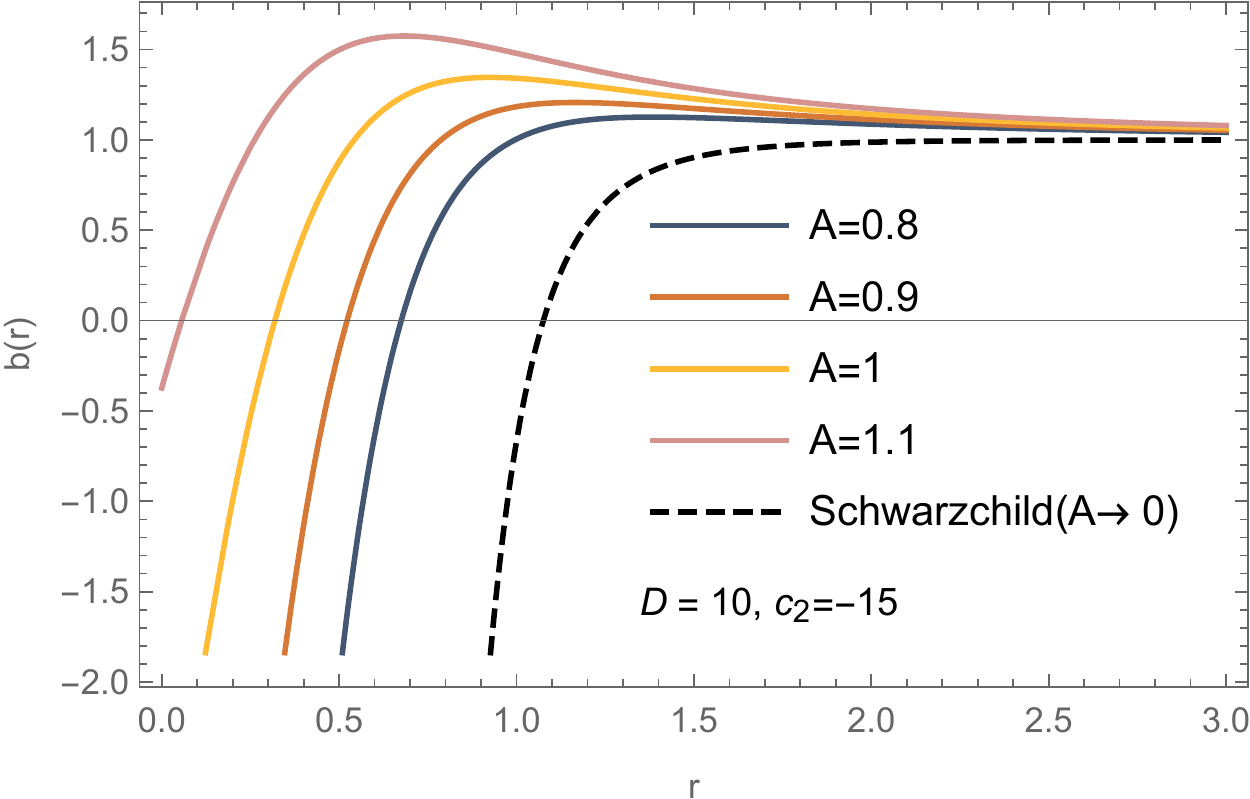}
\caption{The metric function $b(r)$ for asymptotically flat black hole spacetimes from $D=6$ and $D=10$ while varying the mass parameter $c_2$ and the scalar charge.}
\label{d6d10}
\end{figure}

In Fig. \ref{d6d10} we plot the metric functions for the $D=6,10$ cases, where it is clear that for reasonable values of $c_2$, there exists a single horizon. Concluding the discussion for the higher dimensional cases, we have seen that we can obtain compact objects that resemble regular black hole spacetimes with a single horizon. Due to the fact that the spacetime is not asymptotically free from the matter field, we have to define a a cutoff energy scale, above which the theory cannot yield a finite conserved black hole mass and the solutions become void of meaning. Consequently, the higher dimensional black holes do not have a well defined thermodynamical prescription and are not allowed to emit radiation, since their mass is fixed from the theory. Regarding the energy conditions of the black hole, they behave in a similar manner with the lower dimensional cases examined previously.

In the higher dimensional cases, all constants of the solution appear in the potential of the theory, and as a result all constants are fixed, hence nome of them is allowed to vary (in the $D=3$ case we have clearly seen that we are allowed to vary $A$). As a result the only reasonable thermodynamical behavior would be that of an extremal black hole which does not radiate and has a zero temperature, however we are not going to dwell deeper on that matter.

\section{Conclusions}
\label{sec:concl}

In this work we considered $D$-dimensional GR coupled to a phantom scalar field and derived exact regular black hole solutions regardless of the dimensionality of spacetime by examining the behavior of the Kretschmann scalar near the origin. The scalar field possesses a fixed fall-off behavior regardless of the dimension $D$, a feature that has some interesting implications on how the no-hair theorem is violated at higher dimensions. We discussed the cases $D=3,4,5,6,10$ with a detail study of the $D=3,4$ cases.  We investigated the   thermodynamical properties of the solutions  and found that the $D=3$ regular black hole is locally stable, a feature that may be present in AdS black holes due to the growing nature of the temperature. We made  also a detail study of the   $D=4$ black hole which  was also studied in \cite{Bronnikov:2005gm} analysing its thermodynamic properties and we found that it is thermodynamically unstable, since the heat capacity is always negative. We also discussed the properties of the matter threading the black hole. We found that regardless of the dimensionality, the equation of state has a de Sitter nature at the event horizon of the black hole and that the NEC is violated in the causal region of spacetime $r>r_h$.

 Another issue we examined  in this work was the possibility of thermodynamic equilibrium of the compact objects. We have seen that as a result of the generic dependence of the scalar potential on the scalar charge $A$ and mass $m$ of black hole, it is not possible to have thermodynamic equilibrium, and thus the validity of the first law of thermodynamics.
In $D=3,4$ it is possible to eliminate the dependence of the scalar potential from the charge and mass of the object,
given that these potentials depend on the ratio $m/A^{D-1}$, and thus by fixing it, we obtain a secondary hair, and independence of the potential from  individual charge and mass.  The simplicity of the $D=3$ case allowed us to perform simple calculations in order to show that the internal energy of the black hole and the conserved mass do not coincide, and the matter fields influence the internal energy at the event horizon. We have seen that when the scalar hair vanishes the internal energy coincides with the conserved black hole mass. Unfortunately, the $D=4$ case is much more complicated and hence we are not able to show explicitly this behavior, however, there are no signs that this result is dimensional dependant and hence we expect a similar behavior. 

We would like to note here that regular black hole solutions may suffer from the ``mass inflation" instability \cite{massinflation}, which arises because of the unstable nature of the inner horizon. However, our solutions, may not possess an inner horizon (the horizons cannot be obtained analytically, but for reasonable values of the integration constants we can avoid the existence of an inner horizon, while for $D=3$ only a single horizon is present). At least, the existence of an inner horizon is not a necessity as it is for example in the Hayward metric \cite{Hayward:2005gi}. As a result our solutions can avoid the ``mass inflation" instability. However, the stability of the solution against scalar, vector and tensor perturbations is an open problem, which we will leave for future endeavors. Furthermore, in order to integrate the field equations we followed a ``bottom-up" approach and as a result the scalar potential is obtained from the field equations and not fixed a-priori. Hence, it is of interest to include a reasonable scalar potential (as we already discussed, in the presence of a phantom, a positive potential may violate the no-hair theorem, so a combination of even powers in $\phi$ or even a mass term potential may be fruitful to consider), however, due to the difficulty of the field equations, numerical methods should be used.

\section*{Acknowledgements}

The research project was supported by the Hellenic Foundation for Research and Innovation (H.F.R.I.) under the “3rd Call for H.F.R.I. Research Projects to support Post-Doctoral Researchers” (Project Number: 7212). The research of N.E.M. is supported in part by the UK Science and Technology Facilities research Council (STFC) under the research grant ST/T000759/1. N.E.M.  also acknowledges participation in the COST Association Action CA18108 ``{\it Quantum Gravity Phenomenology in the Multimessenger Approach (QG-MM)}''.


\begin{thebibliography}{99}

\bibitem{string} M.~B.~Green, J.~H.~Schwarz and E.~Witten,
``Superstring Theory Vols. 1 and 2:  25th Anniversary Edition,''
(Cambridge University Press, 2012),
ISBN 978-1-139-53477-2, 978-1-107-02911-8
%doi:10.1017/CBO9781139248563 (Vol 1)
and ISBN 978-1-139-53478-9, 978-1-107-02913-2;
%doi:10.1017/CBO9781139248570 (Vol2);
J.~Polchinski,
``String theory. Vol. 1: An introduction to the bosonic string,''
(Cambridge University Press, 2007),
ISBN 978-0-511-25227-3, 978-0-521-67227-6, 978-0-521-63303-1
%doi:10.1017/CBO9780511816079;
``String theory. Vol. 2: Superstring theory and beyond,''
Cambridge University Press, 2007,
ISBN 978-0-511-25228-0, 978-0-521-63304-8, 978-0-521-67228-3.
%doi:10.1017/CBO9780511618123


\bibitem{Kaluza} T. Kaluza,
Sits. Press. Akad. Wiss. Math. Phys. K 1,  895 (1921).

\bibitem{Klein} O. Klein, Z. Phys. K 1, 895, (1921).


%\cite{Mohammedi:2002tv}
\bibitem{Mohammedi:2002tv}
N.~Mohammedi,
``Dynamical compactification, standard cosmology and the accelerating universe,''
Phys. Rev. D \textbf{65}, 104018 (2002)
%doi:10.1103/PhysRevD.65.104018
[arXiv:hep-th/0202119 [hep-th]].
%46 citations counted in INSPIRE as of 12 Nov 2022

%\cite{Ketov:2017aau}
\bibitem{Ketov:2017aau}
S.~V.~Ketov and H.~Nakada,
``Inflation from $(R+\gamma R^n-2\Lambda)$ Gravity in Higher Dimensions,''
Phys. Rev. D \textbf{95}, no.10, 103507 (2017)
%doi:10.1103/PhysRevD.95.103507
[arXiv:1701.08239 [hep-th]].
%15 citations counted in INSPIRE as of 13 Nov 2022

%\cite{Starobinsky:1980te}
\bibitem{Starobinsky:1980te}
A.~A.~Starobinsky,
``A New Type of Isotropic Cosmological Models Without Singularity,''
Phys. Lett. B \textbf{91}, 99-102 (1980)
%10.1016/0370-2693(80)90670-X
%5925 citations counted in INSPIRE as of 13 Nov 2022

%\cite{Keskin:2019mve}
\bibitem{Keskin:2019mve}
A.~I.~Keskin and M.~Salti,
``Cosmographic nature of the early universe from extra dimensional perspective,''
Phys. Lett. B \textbf{791}, 80-85 (2019)
%doi:10.1016/j.physletb.2019.02.024
%4 citations counted in INSPIRE as of 13 Nov 2022

\bibitem{steinhardt}
A.~Upadhye, M.~Ishak and P.~J.~Steinhardt,
``Dynamical dark energy: Current constraints and forecasts,''
Phys. Rev. D \textbf{72} (2005), 063501
%doi:10.1103/PhysRevD.72.063501
[arXiv:astro-ph/0411803 [astro-ph]].


\bibitem{tegmark}
Y.~Wang and M.~Tegmark,
``New dark energy constraints from supernovae, microwave background and galaxy clustering,''
Phys. Rev. Lett. \textbf{92} (2004), 241302
%doi:10.1103/PhysRevLett.92.241302
[arXiv:astro-ph/0403292 [astro-ph]].

\bibitem{seljak}
U.~Seljak \textit{et al.} [SDSS],
``Cosmological parameter analysis including SDSS Ly-alpha forest and galaxy bias: Constraints on the primordial spectrum of fluctuations, neutrino mass, and dark energy,''
Phys. Rev. D \textbf{71} (2005), 103515
%doi:10.1103/PhysRevD.71.103515
[arXiv:astro-ph/0407372 [astro-ph]].

\bibitem{hannestad}
S.~Hannestad and E.~Mortsell,
``Cosmological constraints on the dark energy equation of state and its evolution,''
JCAP \textbf{09} (2004), 001
%doi:10.1088/1475-7516/2004/09/001
[arXiv:astro-ph/0407259 [astro-ph]].

\bibitem{star03}
U.~Alam, V.~Sahni, T.~D.~Saini and A.~A.~Starobinsky,
``Is there supernova evidence for dark energy metamorphosis ?,''
Mon. Not. Roy. Astron. Soc. \textbf{354} (2004), 275
%doi:10.1111/j.1365-2966.2004.08189.x
[arXiv:astro-ph/0311364 [astro-ph]].


\bibitem{chandra}
S.~W.~Allen, R.~W.~Schmidt, H.~Ebeling, A.~C.~Fabian and L.~van Speybroeck,
``Constraints on dark energy from Chandra observations of the largest relaxed galaxy clusters,''
Mon. Not. Roy. Astron. Soc. \textbf{353} (2004), 457
%doi:10.1111/j.1365-2966.2004.08080.x
[arXiv:astro-ph/0405340 [astro-ph]].
%722 citations counted in INSPIRE as of 11 Dec 2022

\bibitem{sen}
A. Sen,
``Rolling tachyon,''
JHEP \textbf{04} (2002), 048
%doi:10.1088/1126-6708/2002/04/048
[arXiv:hep-th/0203211 [hep-th]].
``Tachyon matter,''
JHEP \textbf{07} (2002), 065
%doi:10.1088/1126-6708/2002/07/065
[arXiv:hep-th/0203265 [hep-th]].

\bibitem{gorini}
V.~Gorini, A.~Y.~Kamenshchik, U.~Moschella and V.~Pasquier,
``Tachyons, scalar fields and cosmology,''
Phys. Rev. D \textbf{69} (2004), 123512
%doi:10.1103/PhysRevD.69.123512
[arXiv:hep-th/0311111 [hep-th]].

\bibitem{fara05}
V.~Faraoni,
``Phantom cosmology with general potentials,''
Class. Quant. Grav. \textbf{22} (2005), 3235-3246
%doi:10.1088/0264-9381/22/16/008
[arXiv:gr-qc/0506095 [gr-qc]].

\bibitem{no03}
S.~Nojiri and S.~D.~Odintsov,
``Quantum de Sitter cosmology and phantom matter,''
Phys. Lett. B \textbf{562} (2003), 147-152
%doi:10.1016/S0370-2693(03)00594-X
[arXiv:hep-th/0303117 [hep-th]].

\bibitem{trod}
S.~M.~Carroll, M.~Hoffman and M.~Trodden,
``Can the dark energy equation-of-state parameter $w$ be less than $-1$?,''
Phys. Rev. D \textbf{68} (2003), 023509
%doi:10.1103/PhysRevD.68.023509
[arXiv:astro-ph/0301273 [astro-ph]].

%\cite{Farrah:2023opk}
\bibitem{Farrah:2023opk}
D.~Farrah, K.~S.~Croker, G.~Tarl\'e, V.~Faraoni, S.~Petty, J.~Afonso, N.~Fernandez, K.~A.~Nishimura, C.~Pearson and L.~Wang, \textit{et al.}
``Observational Evidence for Cosmological Coupling of Black Holes and its Implications for an Astrophysical Source of Dark Energy,''
Astrophys. J. Lett. \textbf{944}, no.2, L31 (2023)
%doi:10.3847/2041-8213/acb704
[arXiv:2302.07878 [astro-ph.CO]].
%1 citations counted in INSPIRE as of 23 Feb 2023

%\bibitem{akiyama2022first}
\bibitem[{Akiyama {et~al.}
(2022)Akiyama, Alberdi, Alef, Algaba, Anantua, Asada,
  Azulay, Bach, Baczko, Ball, {et~al.}}]{akiyama2022first}
Akiyama, K., Alberdi, A., Alef, W., {et~al.} 2022, The Astrophysical Journal
  Letters, 930, L12.

%\cite{Bronnikov:2005gm}
\bibitem{Bronnikov:2005gm}
K.~A.~Bronnikov and J.~C.~Fabris,
``Regular phantom black holes,''
Phys. Rev. Lett. \textbf{96} (2006), 251101
%doi:10.1103/PhysRevLett.96.251101
[arXiv:gr-qc/0511109 [gr-qc]].
%266 citations counted in INSPIRE as of 10 Oct 2022
K.~A.~Bronnikov,
``Scalar fields as sources for wormholes and regular black holes,''
Particles \textbf{1} (2018) no.1, 56-81
%doi:10.3390/particles1010005
[arXiv:1802.00098 [gr-qc]].
%42 citations counted in INSPIRE as of 26 Dec 2022

%\cite{Bronnikov:2012ch}
\bibitem{Bronnikov:2012ch}
K.~A.~Bronnikov, R.~A.~Konoplya and A.~Zhidenko,
``Instabilities of wormholes and regular black holes supported by a phantom scalar field,''
Phys. Rev. D \textbf{86} (2012), 024028
%doi:10.1103/PhysRevD.86.024028
[arXiv:1205.2224 [gr-qc]].
%127 citations counted in INSPIRE as of 26 Dec 2022


\bibitem{dym92}
I.~Dymnikova,
``Vacuum nonsingular black hole,''
Gen. Rel. Grav. \textbf{24}, 235-242 (1992).
%doi:10.1007/BF00760226
%418 citations counted in INSPIRE as of 14 Nov 2022

\bibitem{ned01}
K.~A.~Bronnikov,
``Regular magnetic black holes and monopoles from nonlinear electrodynamics,''
Phys. Rev. D \textbf{63} (2001), 044005
%doi:10.1103/PhysRevD.63.044005
[arXiv:gr-qc/0006014 [gr-qc]].

\bibitem{bdd03}
K.~A.~Bronnikov, A.~Dobosz and I.~G.~Dymnikova,
``Nonsingular vacuum cosmologies with a variable cosmological term,''
Class. Quant. Grav. \textbf{20} (2003), 3797-3814
%doi:10.1088/0264-9381/20/16/317
[arXiv:gr-qc/0302029 [gr-qc]].

%\cite{Ma:2014qma}
\bibitem{Ma:2014qma}
M.~S.~Ma and R.~Zhao,
``Corrected form of the first law of thermodynamics for regular black holes,''
Class. Quant. Grav. \textbf{31} (2014), 245014
%doi:10.1088/0264-9381/31/24/245014
[arXiv:1411.0833 [gr-qc]].
%51 citations counted in INSPIRE as of 10 Apr 2023

%\cite{Lan:2023cvz}
\bibitem{Lan:2023cvz}
C.~Lan, H.~Yang, Y.~Guo and Y.~G.~Miao,
``Regular black holes: A short topic review,''
[arXiv:2303.11696 [gr-qc]].
%1 citations counted in INSPIRE as of 10 Apr 2023


\bibitem{primary}  D.~C.~Zou and Y.~S.~Myung,
``Black hole with primary scalar hair in Einstein-Weyl-Maxwell-conformal scalar theory,''
Phys. Rev. D \textbf{101}, no.8, 084021 (2020)
%doi:10.1103/PhysRevD.101.084021
[arXiv:2001.01351 [gr-qc]];

A.~Anabalon, F.~Canfora, A.~Giacomini and J.~Oliva,
``Black Holes with Primary Hair in gauged N=8 Supergravity,''
JHEP \textbf{06}, 010 (2012)
%doi:10.1007/JHEP06(2012)010
[arXiv:1203.6627 [hep-th]];

S.~Mignemi and D.~L.~Wiltshire,
``Multi-scalar black holes with contingent primary hair: Mechanics and stability,''
Phys. Rev. D \textbf{70}, 124012 (2004)
%doi:10.1103/PhysRevD.70.124012
[arXiv:hep-th/0408215 [hep-th]];

S.~Mignemi,
``Primary scalar hair in dilatonic theories with modulus fields,''
Phys. Rev. D \textbf{62}, 024014 (2000)
%doi:10.1103/PhysRevD.62.024014
[arXiv:gr-qc/9910041 [gr-qc]];

P.~A.~Gonz\'alez, E.~Papantonopoulos, J.~Saavedra and Y.~V\'asquez,
``Four-Dimensional Asymptotically AdS Black Holes with Scalar Hair,''
JHEP \textbf{12} (2013), 021
%doi:10.1007/JHEP12(2013)021
[arXiv:1309.2161 [gr-qc]].

%\cite{Jacobson:2007tj}
\bibitem{Jacobson:2007tj}
T.~Jacobson,
``When is g(tt) g(rr) = -1?,''
Class. Quant. Grav. \textbf{24} (2007), 5717-5719
%doi:10.1088/0264-9381/24/22/N02
[arXiv:0707.3222 [gr-qc]].
%100 citations counted in INSPIRE as of 24 Dec 2022

%\cite{Faraoni:2021gdl}
\bibitem{Faraoni:2021gdl}
V.~Faraoni and A.~Leblanc,
``Disformal mappings of spherical DHOST geometries,''
JCAP \textbf{08} (2021), 037
%doi:10.1088/1475-7516/2021/08/037
[arXiv:2107.03456 [gr-qc]].
%9 citations counted in INSPIRE as of 25 Dec 2022

%\cite{Chatzifotis:2021hpg}
\bibitem{Chatzifotis:2021hpg}
N.~Chatzifotis, E.~Papantonopoulos and C.~Vlachos,
``Disformal transition of a black hole to a wormhole in scalar-tensor Horndeski theory,''
Phys. Rev. D \textbf{105} (2022) no.6, 064025
%doi:10.1103/PhysRevD.105.064025
[arXiv:2111.08773 [gr-qc]].
%8 citations counted in INSPIRE as of 25 Dec 2022

%\cite{Bronnikov:2022ofk}
\bibitem{Bronnikov:2022ofk}
K.~A.~Bronnikov,
``Regular black holes sourced by nonlinear electrodynamics,''
[arXiv:2211.00743 [gr-qc]].
%2 citations counted in INSPIRE as of 12 Apr 2023



%\cite{Bekenstein:1972ny}
\bibitem{Bekenstein:1972ny}
J.~D.~Bekenstein,
``Transcendence of the law of baryon-number conservation in black hole physics,''
Phys. Rev. Lett. \textbf{28} (1972), 452-455;
%doi:10.1103/PhysRevLett.28.452
%286 citations counted in INSPIRE as of 01 Jan 2023
%\cite{Bekenstein:1971hc}
J.~D.~Bekenstein,
``Nonexistence of baryon number for static black holes,''
Phys. Rev. D \textbf{5} (1972), 1239-1246.
%doi:10.1103/PhysRevD.5.1239
%521 citations counted in INSPIRE as of 01 Jan 2023

%\cite{Farakos:2009fx}
\bibitem{Farakos:2009fx}
K.~Farakos, A.~P.~Kouretsis and P.~Pasipoularides,
``Anti de Sitter 5D black hole solutions with a self-interacting bulk scalar field: A Potential reconstruction approach,''
Phys. Rev. D \textbf{80}, 064020 (2009)
%doi:10.1103/PhysRevD.80.064020
[arXiv:0905.1345 [hep-th]].
%8 citations counted in INSPIRE as of 28 Oct 2022

%\cite{potential}
\bibitem{potential}
%\cite{Herdeiro:2015waa}
C.~A.~R.~Herdeiro and E.~Radu,
``Asymptotically flat black holes with scalar hair: a review,''
Int. J. Mod. Phys. D \textbf{24} (2015) no.09, 1542014
%doi:10.1142/S0218271815420146
[arXiv:1504.08209 [gr-qc]];
%496 citations counted in INSPIRE as of 02 Jan 2023

A.~Bakopoulos and T.~Nakas,
``Analytic and asymptotically flat hairy (ultra-compact) black-hole solutions and their axial perturbations,''
JHEP \textbf{04} (2022), 096
%doi:10.1007/JHEP04(2022)096
[arXiv:2107.05656 [gr-qc]];
%4 citations counted in INSPIRE as of 02 Jan 2023

%\cite{Karakasis:2022xzm}
T.~Karakasis, G.~Koutsoumbas, A.~Machattou and E.~Papantonopoulos,
``Magnetically charged Euler-Heisenberg black holes with scalar hair,''
Phys. Rev. D \textbf{106} (2022) no.10, 104006
%doi:10.1103/PhysRevD.106.104006
[arXiv:2207.13146 [gr-qc]].
%0 citations counted in INSPIRE as of 02 Jan 2023

%\cite{Bakopoulos:2023hkh}
A.~Bakopoulos and T.~Nakas,
``Novel exact ultra-compact and ultra-sparse hairy black holes emanating from regular and phantom scalar fields,''
[arXiv:2303.09116 [gr-qc]].
%0 citations counted in INSPIRE as of 11 Apr 2023

%\cite{Chew:2022enh}
X.~Y.~Chew, D.~h.~Yeom and J.~L.~Bl\'azquez-Salcedo,
``Properties of Scalar Hairy Black Holes and Scalarons with Asymmetric Potential,''
[arXiv:2210.01313 [gr-qc]].
%1 citations counted in INSPIRE as of 04 May 2023

%\cite{Cataldo:2000ns}
\bibitem{regular3}
M.~Cataldo and A.~Garcia,
``Regular (2+1)-dimensional black holes within nonlinear electrodynamics,''
Phys. Rev. D \textbf{61} (2000), 084003
%doi:10.1103/PhysRevD.61.084003
[arXiv:hep-th/0004177 [hep-th]];
%70 citations counted in INSPIRE as of 01 Jan 2023

%\cite{HabibMazharimousavi:2011gh}
S.~Habib Mazharimousavi, M.~Halilsoy and T.~Tahamtan,
``Regular charged black hole construction in 2+1 -dimensions,''
Phys. Lett. A \textbf{376} (2012), 893-898
%doi:10.1016/j.physleta.2012.01.001
[arXiv:1107.0242 [gr-qc]];
%14 citations counted in INSPIRE as of 01 Jan 2023

%\cite{Jusufi:2023fpo}
K.~Jusufi, M.~Jamil and A.~Sheykhi,
``Regular 3D Gauss-Bonnet black holes with finite electrodynamics,''
[arXiv:2302.10799 [physics.gen-ph]];
%0 citations counted in INSPIRE as of 22 Feb 2023

%\cite{Estrada:2020tbz}
M.~Estrada and F.~Tello-Ortiz,
``A new model of regular black hole in (2+1) dimensions,''
EPL \textbf{135} (2021) no.2, 20001
%doi:10.1209/0295-5075/ac0ed0
[arXiv:2012.05068 [gr-qc]];
%8 citations counted in INSPIRE as of 01 Jan 2023

%\cite{Bueno:2021krl}
P.~Bueno, P.~A.~Cano, J.~Moreno and G.~van der Velde,
``Regular black holes in three dimensions,''
Phys. Rev. D \textbf{104} (2021) no.2, L021501
%doi:10.1103/PhysRevD.104.L021501
[arXiv:2104.10172 [gr-qc]];
%22 citations counted in INSPIRE as of 01 Jan 2023

%\cite{Bueno:2022ewf}
P.~Bueno, P.~A.~Cano, J.~Moreno and G.~van der Velde,
``Electromagnetic Generalized Quasi-topological gravities in $(2 + 1)$ dimensions,''
[arXiv:2212.00637 [gr-qc]].
%1 citations counted in INSPIRE as of 01 Jan 2023

\bibitem{BTZ}
M.~Banados, C.~Teitelboim and J.~Zanelli,
``The Black hole in three-dimensional space-time,''
Phys. Rev. Lett. \textbf{69} (1992), 1849-1851
%doi:10.1103/PhysRevLett.69.1849
[arXiv:hep-th/9204099 [hep-th]];

M.~Banados, M.~Henneaux, C.~Teitelboim and J.~Zanelli,
``Geometry of the (2+1) black hole,''
Phys. Rev. D \textbf{48} (1993), 1506-1525
[erratum: Phys. Rev. D \textbf{88} (2013), 069902]
%doi:10.1103/PhysRevD.48.1506
[arXiv:gr-qc/9302012 [gr-qc]];

S.~Carlip,
``The (2+1)-Dimensional black hole,''
Class. Quant. Grav. \textbf{12} (1995), 2853-2880
%doi:10.1088/0264-9381/12/12/005
[arXiv:gr-qc/9506079 [gr-qc]].

%\cite{Chan:1995fr}
\bibitem{quasilocal}
K.~C.~K.~Chan, J.~H.~Horne and R.~B.~Mann,
``Charged dilaton black holes with unusual asymptotics,''
Nucl. Phys. B \textbf{447} (1995), 441-464
%doi:10.1016/0550-3213(95)00205-7
[arXiv:gr-qc/9502042 [gr-qc]];
%236 citations counted in INSPIRE as of 04 Jan 2023

J.~D.~Brown and J.~W.~York, Jr.,
``Quasilocal energy and conserved charges derived from the gravitational action,''
Phys. Rev. D \textbf{47} (1993), 1407-1419
%doi:10.1103/PhysRevD.47.1407
[arXiv:gr-qc/9209012 [gr-qc]].
%1482 citations counted in INSPIRE as of 04 Jan 2023

%\cite{Wald:1993nt}
\bibitem{Wald:1993nt}
R.~M.~Wald,
``Black hole entropy is the Noether charge,''
Phys. Rev. D \textbf{48} (1993) no.8, R3427-R3431
%doi:10.1103/PhysRevD.48.R3427
[arXiv:gr-qc/9307038 [gr-qc]].
%1988 citations counted in INSPIRE as of 07 Dec 2022


%\cite{singularbtz}
\bibitem{singularbtz}
F.~Correa, A.~Fa\'undez and C.~Mart\'\i{}nez,
``Rotating hairy black hole and its microscopic entropy in three spacetime dimensions,''
Phys. Rev. D \textbf{87} (2013) no.2, 027502
%doi:10.1103/PhysRevD.87.027502
[arXiv:1211.4878 [hep-th]];

K.~C.~K.~Chan,
``Modifications of the BTZ black hole by a dilaton / scalar,''
Phys. Rev. D \textbf{55} (1997), 3564-3574
%doi:10.1103/PhysRevD.55.3564
[arXiv:gr-qc/9603038 [gr-qc]];

T.~Karakasis, E.~Papantonopoulos, Z.~Y.~Tang and B.~Wang,
``Black holes of (2+1)-dimensional $f(R)$ gravity coupled to a scalar field,''
Phys. Rev. D \textbf{103} (2021) no.6, 064063
%doi:10.1103/PhysRevD.103.064063
[arXiv:2101.06410 [gr-qc]].
%16 citations counted in INSPIRE as of 19 Jan 2023

%\cite{Karakasis:2021ttn}
T.~Karakasis, E.~Papantonopoulos, Z.~Y.~Tang and B.~Wang,
``(2+1)-dimensional black holes in f(R,\ensuremath{\phi}) gravity,''
Phys. Rev. D \textbf{105} (2022) no.4, 044038
%doi:10.1103/PhysRevD.105.044038
[arXiv:2201.00035 [gr-qc]].
%7 citations counted in INSPIRE as of 11 Apr 2023

%\cite{Karakasis:2022fep}
T.~Karakasis, E.~Papantonopoulos, Z.~Y.~Tang and B.~Wang,
``Rotating (2+1)-dimensional black holes in Einstein-Maxwell-dilaton theory,''
Phys. Rev. D \textbf{107} (2023) no.2, 024043
%doi:10.1103/PhysRevD.107.024043
[arXiv:2210.15704 [gr-qc]].
%3 citations counted in INSPIRE as of 11 Apr 2023

%\cite{Reznik:1994py}
\bibitem{Reznik:1994py}
B.~Reznik,
``Thermodynamics and evaporation of the (2+1)-dimensions black hole,''
Phys. Rev. D \textbf{51} (1995), 1728-1732
%doi:10.1103/PhysRevD.51.1728
[arXiv:gr-qc/9403027 [gr-qc]].
%20 citations counted in INSPIRE as of 26 Jan 2023

%\cite{Ditta:2020jud}
\bibitem{Ditta:2020jud}
A.~Ditta and G.~Abbas,
``Circular orbits and accretion process near a regular phantom black hole,''
Gen. Rel. Grav. \textbf{52} (2020) no.8, 77
%doi:10.1007/s10714-020-02724-9
%7 citations counted in INSPIRE as of 12 Apr 2023

%\cite{Ding:2013vta}
\bibitem{Ding:2013vta}
C.~Ding, C.~Liu, Y.~Xiao, L.~Jiang and R.~G.~Cai,
``Strong gravitational lensing in a black-hole spacetime dominated by dark energy,''
Phys. Rev. D \textbf{88} (2013) no.10, 104007
%doi:10.1103/PhysRevD.88.109907
[arXiv:1308.5035 [gr-qc]].
%44 citations counted in INSPIRE as of 12 Apr 2023

%\cite{Altamirano:2014tva}
\bibitem{Altamirano:2014tva}
N.~Altamirano, D.~Kubiznak, R.~B.~Mann and Z.~Sherkatghanad,
``Thermodynamics of rotating black holes and black rings: phase transitions and thermodynamic volume,''
Galaxies \textbf{2} (2014), 89-159
%doi:10.3390/galaxies2010089
[arXiv:1401.2586 [hep-th]].
%285 citations counted in INSPIRE as of 28 Dec 2022


%\cite{Tangherlini:1963bw}
\bibitem{Tangherlini:1963bw}
F.~R.~Tangherlini,
``Schwarzschild field in n dimensions and the dimensionality of space problem,''
Nuovo Cim. \textbf{27} (1963), 636-651
%doi:10.1007/BF02784569
%753 citations counted in INSPIRE as of 31 Dec 2022

%\cite{Poisson:1989zz}
\bibitem{massinflation}
E.~Poisson and W.~Israel,
``Inner-horizon instability and mass inflation in black holes,''
Phys. Rev. Lett. \textbf{63} (1989), 1663-1666:
%doi:10.1103/PhysRevLett.63.1663
%210 citations counted in INSPIRE as of 04 Jan 2023

%\cite{Carballo-Rubio:2018pmi}
R.~Carballo-Rubio, F.~Di Filippo, S.~Liberati, C.~Pacilio and M.~Visser,
``On the viability of regular black holes,''
JHEP \textbf{07} (2018), 023
%doi:10.1007/JHEP07(2018)023
[arXiv:1805.02675 [gr-qc]];
%91 citations counted in INSPIRE as of 04 Jan 2023

%\cite{Carballo-Rubio:2022kad}
R.~Carballo-Rubio, F.~Di Filippo, S.~Liberati, C.~Pacilio and M.~Visser,
``Regular black holes without mass inflation instability,''
JHEP \textbf{09} (2022), 118
%doi:10.1007/JHEP09(2022)118
[arXiv:2205.13556 [gr-qc]].
%12 citations counted in INSPIRE as of 04 Jan 2023

%\cite{Hayward:2005gi}
\bibitem{Hayward:2005gi}
S.~A.~Hayward,
``Formation and evaporation of regular black holes,''
Phys. Rev. Lett. \textbf{96} (2006), 031103
%doi:10.1103/PhysRevLett.96.031103
[arXiv:gr-qc/0506126 [gr-qc]].
%748 citations counted in INSPIRE as of 04 Jan 2023

%\cite{Cardenas:2022jtz}
%\bibitem{Cardenas:2022jtz}
%M.~C\'ardenas, O.~Fuentealba, C.~Mart\'\i{}nez and R.~Troncoso,
%``Gravity coupled to a scalar field from a Chern-Simons action: describing rotating hairy black holes and solitons with gauge fields,''
%[arXiv:2212.13094 [hep-th]].
%0 citations counted in INSPIRE as of 09 Jan 2023



%\cite{Feng:2013tza}
\bibitem{Feng:2013tza}
X.~H.~Feng, H.~Lu and Q.~Wen,
``Scalar Hairy Black Holes in General Dimensions,''
Phys. Rev. D \textbf{89} (2014) no.4, 044014
%doi:10.1103/PhysRevD.89.044014
[arXiv:1312.5374 [hep-th]].
%52 citations counted in INSPIRE as of 05 Feb 2023


\end{thebibliography}
\end{document}